\documentclass[english, 9 pt]{paper}
\usepackage[T1]{fontenc}
\usepackage[latin9]{inputenc}
\usepackage{geometry}
\usepackage{amsmath}
\usepackage{graphicx}
\usepackage{setspace}

\usepackage{url}
\usepackage{hyperref}

\usepackage{natbib} \bibpunct{(}{)}{;}{author-year}{}{,}

\usepackage[explicit]{titlesec}
\titleformat{\section}
   {\centering\large\sffamily\bfseries\sc}
   {\thesection.}
   {0.5em}
   {#1}

\usepackage[format=plain,%
             labelfont={bf,sf},%
             labelsep=space,%
             figurename=Figure]{caption}

\usepackage{fancyhdr}

\usepackage{lscape}
\usepackage{fancyhdr}
\usepackage[titletoc,title]{appendix}

\makeatletter

\providecommand{\tabularnewline}{\\}
\newcommand{\lyxdot}{.}

\@ifundefined{showcaptionsetup}{}{%
 \PassOptionsToPackage{caption=false}{subfig}}
\usepackage{subfig}
\makeatother

\usepackage{babel}
\begin{document}
\global\long\def\E{\mathrm{E}}

\title{Bayesian co-estimation of selfing rate and locus-specific mutation
rates for a partially selfing population}

\author{Benjamin D. Redelings, Seiji Kumagai, Liuyang Wang,\\
Andrey Tatarenkov, Ann K. Sakai, Stephen G. Weller,\\
Theresa M. Culley, John C. Avise, and Marcy K. Uyenoyama}
\maketitle

\section*{Abstract}

We present a Bayesian method for characterizing the mating system of populations reproducing through a mixture of self-fertilization and random outcrossing.  
Our method uses patterns of genetic variation across the genome as a basis for inference about pure hermaphroditism, androdioecy, and gynodioecy.  We extend the standard coalescence model to accommodate these mating systems, accounting explicitly for multilocus identity disequilibrium, inbreeding depression, and variation in fertility among mating types.  We incorporate the Ewens Sampling Formula (ESF) under the infinite-alleles model of mutation to obtain a novel expression for the likelihood of mating system parameters.  Our Markov chain
Monte Carlo (MCMC) algorithm assigns locus-specific mutation rates, drawn from a common mutation rate distribution that is itself estimated from the data using a Dirichlet Process Prior model.  Among the parameters jointly inferred are the population-wide rate of self-fertilization, locus-specific mutation rates, and the number of generations since the most recent outcrossing event for each sampled individual.

\section{Introduction}
Inbreeding has pervasive consequences throughout the genome, affecting genealogical relationships between genes held at each locus within individuals and among multiple loci.  This generation of genome-wide, multilocus disequilibria of various orders transforms the context in which evolution proceeds.  Here, we address a simple form of inbreeding:  a mixture of self-fertilization (selfing) and random outcrossing \citep{Clegg1980, Ritland2002}.

Various methods exist for the estimation of selfing rates from genetic data.  Wright's (\citeyear{Wright1921}) fundamental approach bases the estimation of selfing rates on the coefficient of inbreeding ($F_{IS}$), which reflects the departure from Hardy-Weinberg proportions of genotypes for a given set of allele frequencies.  The maximum likelihood method of \citet{Enjalbert2000} detects inbreeding from departures of multiple loci from Hardy-Weinberg proportions, estimating allele frequencies for each locus and accounting for correlations in heterozygosity among loci \citep[identity disequilibrium,][]{Cockerham1968}.  \citet{David2007} extend the approach of \citet{Enjalbert2000}, basing the estimation of selfing rates on the distribution of heterozygotes across multiple, unlinked loci, while accommodating errors in scoring heterozygotes as homozygotes.  A primary objective of \texttt{InStruct} \citep{Gao2007} is the estimation of admixture.  It extends the widely-used program \texttt{structure} \citep{Pritchard2000}, which bases the estimation of admixture on disequilibria of various forms, by accounting for disequilibria due to selfing.  Progeny array methods \citep[see][]{Ritland2002}, which base the estimation of selfing rates on the genetic analysis of progeny for which one or more parents are known, are particularly well-suited to plant populations.  \citet{Wang2012} extend this approach to a random sample of individuals by reconstructing sibship relationships within the sample.

Methods that base the estimation of inbreeding rates on the observed departure from random union of gametes require information on expected Hardy-Weinberg proportions.  Population-wide frequencies of alleles observed in a sample at locus $l$ ($\{p_{li}\}$) 
can be estimated jointly in a maximum-likelihood framework \citep[\emph{e.g.},][]{HillWalliker1995} or integrated out as nuisance parameters in a Bayesian framework \citep[\emph{e.g.},][]{AyresBalding1998}.  Similarly,  locus-specific heterozygosity
\begin{equation}
\label{apphet}
d_l=1-\sum_i p_{li}^2
\end{equation}
can be obtained from observed allele frequencies \citep{Enjalbert2000} or estimated directly and jointly with the selfing rate \citep{David2007}.

In contrast, our Bayesian method for the analysis of partial self-fertilization derives from a coalescence model that accounts for genetic variation and uses the Ewens Sampling Formula \citep[ESF,][]{Ewens1972}.  Our approach replaces the estimation of allele frequencies or heterozygosity \eqref{apphet} by the estimation of a locus-specific mutation rate ($\theta^\ast$) under the infinite-alleles model of mutation. We use a Dirichlet Process Prior (DPP) to determine the number of classes of mutation rates, the mutation rate for each class, and the class membership of each locus.  We assign the DPP parameters in a conservative manner so that it creates a new mutational class only if sufficient evidence exists to justify doing so.  Further, while other methods assume that the frequency in the population of an allelic class not observed in the sample is zero, the ESF provides the probability, under the infinite-alleles model of mutation, that the next-sampled gene represents a novel allele (see \eqref{novelpi}).

To estimate the probability that a random individual is uniparental ($s^\ast$),
we exploit identity disequilibrium \citep{Cockerham1968}, the correlation in heterozygosity across loci.  This association, even among unlinked loci, reflects that all loci within an individual share a history of inbreeding back to the most recent random outcrossing event.  Conditional on the number of generations since this event, the genealogical histories of unlinked loci are independent.  Our method infers the number of consecutive generations of self-fertilization in the immediate ancestry of each sampled diploid individual and the probability of coalescence during this period between the lineages at each locus.

In inferring the full likelihood from the observed frequency spectrum of diploid genotypes at multiple unlinked loci, we determine the distributions of the allele frequency spectra ancestral to the sample at the most recent point at which all sampled gene lineages at each locus reside in separate individuals.  At this point, the ESF provides the exact likelihood, obviating the need for further genealogical reconstruction.  This approach permits computationally efficient analysis of samples comprising large numbers of individuals and large numbers of loci observed across the genome.

Here, we address the estimation of inbreeding rates in populations undergoing pure hermaphroditism, androdioecy (hermaphrodites and males), or gynodioecy (hermaphrodites and females).  Our method provides a means for the simultaneous inference of various aspects of the mating system, including the population proportions of sexual forms and levels of inbreeding depression.
We apply our method to simulated data sets to demonstrate its accuracy in parameter estimation and in  assessing uncertainty.
Our application to microsatellite data from the androdioecious killifish \emph{Kryptolebias marmoratus} \citep{Mackiewicz2006b, TatarenkovEarleyTaylorAvise2012} and to the gynodioecious Hawaiian endemic \emph{Schiedea salicaria} \citep{WallaceCulleyWellerSakaiNepopkroeff2011} illustrates the formation of inferences about a number of biologically significant aspects, including measures of effective population size.

\section*{Evolutionary model}

We describe our use of the Ewens Sampling Formula \citep[ESF,][]{Ewens1972} to determine likelihoods based on a sample of diploid multilocus genotypes. 

From a reduced sample, formed by subsampling a single gene from each locus from each diploid individual, one could use the ESF to determine a likelihood function with a single parameter:  the mutation rate, appropriately scaled to account for the acceleration of the coalescence rate caused by inbreeding \citep{Nordborg1997, Fu1997}.  Observation of diploid genotypes provides information about another parameter:  the probability that a random individual is uniparental (uniparental proportion).   We describe the dependence of these two composite parameters on the basic parameters of models of pure hermaphroditism, androdioecy, and gynodioecy.

\subsection*{Rates of coalescence and mutation}\label{sec:coalmut}

Here, we describe the structure of the coalescence process shared by our models of pure hermaphroditism, androdioecy, and gynodioecy.

\paragraph{Relative rates of coalescence and mutation:}  We represent the probability that a random individual is uniparental by $s^\ast$ and the probability that a pair of genes that reside in distinct individuals descend from the same parent in the immediately preceding generation by $1/N^\ast$.  These quantities determine the coalescence rate and the scaled mutation rate of the ESF.

A pair of lineages residing in distinct individuals derive from a single parent (P) in the preceding generation at rate $1/N^\ast$.  They descend from the same gene (immediate coalescence) or from distinct genes in that individual with equal probability.  In the latter case, P is either uniparental (probability $s^\ast$), implying descent once again of the lineages from a single individual in the preceding generation, or biparental, implying descent from distinct individuals.  Residence of a pair of lineages in a single individual rapidly resolves either to coalescence, with probability
\begin{equation}
\label{pcoal}
f_c=\frac{s^\ast}{2-s^\ast},
\end{equation}
or to residence in distinct individuals, with the complement probability.  This expression is identical to the classical coefficient of identity \citep{Wright1921, Haldane1924}.
The total rate of coalescence of lineages sampled from distinct individuals corresponds to
\begin{equation}
\label{coalrate}
\frac{(1+f_c)/2}{N^\ast} =\frac{1}{N^\ast(2-s^\ast)}.
\end{equation}

Our model assumes that coalescence and mutation occur on comparable time scales:
\begin{equation}
\begin{split}
\label{rates}
\lim_{\substack{
N \to \infty\\
u \to 0
}} 4Nu &= \theta\\
\lim_{\substack{
N \to \infty\\
N^\ast \to \infty
}} N^\ast/N &= S,
\end{split}
\end{equation}
for $u$ the rate of mutation under the infinite alleles model and $N$ an arbitrary quantity that goes to infinity at a rate comparable to $N^\ast$ and $1/u$.  Here, $S$ represents a scaled measure of effective population size \citep[termed ``inbreeding effective size'' by][]{CrowDenniston88}, relative to a population comprising $N$ reproductives.

In large populations, switching of lineages between uniparental and biparental carriers occurs on the order of generations, virtually instantaneously relative to the rate at which lineages residing in distinct individuals coalesce \citep{Nordborg1997, Fu1997}.  Our model assumes independence between the processes of coalescence and mutation and that these processes occur on a much longer time scale than 
random outcrossing:
\begin{equation}
\label{eq:assumptions}
1 - s^\ast \gg u, 1/N^\ast.
\end{equation}
For $m$ lineages, each residing in a distinct individual, the probability that the most recent event corresponds to mutation is
\begin{equation*}
\lim_{N \to \infty} \frac{mu}{mu+\binom{m}{2}/[N^\ast(2-s^\ast)]}=\frac{\theta^\ast}{\theta^\ast+m-1},
\end{equation*}
in which
\begin{align}
\label{eq:theta-star}
\theta^\ast &= 
\lim_{\substack{
N \to \infty\\
u \to 0
}}
2N^\ast u(2-s^\ast) = \lim_{\substack{
N \to \infty\\
u \to 0
}}4Nu \; \frac{N^\ast}{N} (1-s^\ast /2) \nonumber \\
&= \theta (1-s^\ast /2)S,
\end{align}
for $\theta$ and $S$ defined in \eqref{rates}.  In inbred populations, the single parameter of the ESF corresponds to $\theta^\ast$.

\paragraph{Uniparental proportion and the rate of parent-sharing:}  In a population comprising $N_h$ hermaphrodites, the rate of parent-sharing corresponds to $1/N_h$, and the uniparental proportion ($s^\ast$) corresponds to
\begin{subequations}
\label{eq:pure-h}
\begin{equation}
s_{H}= \frac{\tilde{s}\tau}{\tilde{s}\tau+1-\tilde{s}},
\end{equation}
for $\tilde{s}$ the fraction of uniparental offspring at conception and
$\tau$ the rate of survival
of uniparental relative to biparental offspring.  For the pure-hermaphroditism model, we assign the arbitrary constant $N$ in \eqref{rates} as $N_{h}$, implying
\begin{equation}
S_H \equiv 1.
\end{equation}
\end{subequations}

In androdioecious populations, comprising $N_{h}$ reproducing hermaphrodites and
$N_{m}$ reproducing males (female-steriles),
the uniparental proportion ($s^\ast$)
is identical to the case of pure hermaphroditism \eqref{eq:pure-h}
\begin{subequations}
\label{eq:androdioecy}
\begin{equation}
\label{eq:andros}
s_{A}= \frac{\tilde{s}\tau}{\tilde{s}\tau+1-\tilde{s}}.
\end{equation}
A random gene derives from a male in the preceding generation with probability
\begin{equation*}
(1-s_{A})/2,
\end{equation*}
and from a hermaphrodite with the complement probability.  A pair of genes sampled from distinct individuals derive from the same parent ($1/N^{*}$) with probability
\begin{equation}
\label{parsharea}
\frac{1}{N_{A}} = \frac{[(1+s_{A})/2]^2}{N_{h}}+\frac{[(1-s_{A})/2]^2}{N_{m}}.
\end{equation}
In the absence of inbreeding ($s_{A}=0$), this
expression agrees with the classical harmonic mean expression for effective
population size \citep{Wright69}.  For the androdioecy model, we assign the arbitrary constant in \eqref{rates} as the number of reproductives $(N_{h}+N_{m})$,
implying a scaled rate of coalescence corresponding to
\begin{equation}
\frac{1}{S_A}= \frac{N_{h}+N_{m}}{N_{A}}= \frac{[(1+s_{A})/2]^{2}}{1-p_{m}}+\frac{[(1-s_{A})/2]^{2}}{p_{m}},
\end{equation}
\end{subequations}
for
\begin{equation}
\label{definerhom}
p_{m}=\frac{N_{m}}{N_{h}+N_{m}}
\end{equation}
the proportion of males among reproductive individuals. Relative effective number $S_A \in (0,1]$ takes its maximum for populations in which the effective number $N_A$, implied by the rate of parent sharing, corresponds to the total number of reproductives ($N_A=N_{h}+N_{m}$).  At $S_A =1$, the probability that a random gene derives from a male parent equals the proportion of males among reproductives:
\begin{equation*}
(1-s_{A})/2 =p_m.
\end{equation*}

In gynodioecious populations, in which $N_{h}$ hermaphrodites and
$N_{f}$ females (male-steriles) reproduce, the uniparental proportion ($s^\ast$)
corresponds to
\begin{subequations}
\label{eq:gynodioecy}
\begin{equation}
\label{eq:gynos}
s_{G} = \frac{\tau N_{h}a}{\tau N_{h}a+N_{h}(1-a)+N_{f}\sigma},
\end{equation}
in which $\sigma$ represents the seed fertility of females relative to 
hermaphrodites and $a$ the proportion of seeds of hermaphrodites set by self-pollen.  A random gene derives from a female in the preceding generation with probability
\begin{equation*}
(1-s_{G})F/2,
\end{equation*}
for
\begin{equation}
\label{eq:H}
F = \frac{N_{f}\sigma}{N_{h}(1-a)+N_{f}\sigma}
\end{equation}
the proportion of biparental offspring that have a female parent.  A pair of genes sampled from distinct individuals derive from the same parent ($1/N^{*}$) with probability
\begin{equation}
\label{parshareg}
\frac{1}{N_{G}}  = \frac{[1-(1-s_{G})F/2]^{2}}{N_{h}}+\frac{[(1-s_{G})F/2]^{2}}{N_{f}}.
\end{equation}
We assign the arbitrary constant $N$ in \eqref{rates} as $(N_{h}+N_{f})$, implying a scaled rate of coalescence of
\begin{equation}
\frac{1}{S_G}=  \frac{N_{h}+N_{f}}{N_{G}}=\frac{[1-(1-s_{G})F/2]^{2}}{1-p_{f}}+\frac{[(1-s_{G})F/2]^{2}}{p_{f}},
\end{equation}
\end{subequations}
for
\begin{equation}
\label{definerhof}
p_{f}=\frac{N_{f}}{N_{h}+N_{f}}
\end{equation}
the proportion of females  among reproductive individuals.  As for the androdioecy model, $S_G \in (0,1]$ achieves its maximum only if the proportion of females among reproductives equals the probability that a random gene derives from a female parent:
\begin{equation*}
(1-s_{G})F/2 =p_f.
\end{equation*}

\subsection*{Likelihood}

We here address the probability of a sample of diploid multilocus genotypes.

\paragraph{Genealogical histories:} For a sample comprising up to two alleles at each of $L$ autosomal loci in $n$ diploid individuals, we represent the observed genotypes by 
\begin{equation}
\label{eq:X}
\mathbf{X} = \left\{ \mathbf{X}_{1},\mathbf{X}_{2},\ldots,\mathbf{X}_{L}\right\},
\end{equation}
in which $\mathbf{X}_{l}$ denotes the set of genotypes observed at locus
$l$,
\begin{equation}
\label{eq:X_l}
\mathbf{X}_{l} = \left\{ \mathbf{X}_{l1,}\mathbf{X}_{l2},\ldots,\mathbf{X}_{ln}\right\} ,
\end{equation}
with
\begin{eqnarray*}
\mathbf{X}_{lk} & = & (X_{lk1},X_{lk2})
\end{eqnarray*}
the genotype at locus $l$ of individual $k$, with alleles $X_{lk1}$
and $X_{lk2}$.

To facilitate accounting for the shared recent history of genes borne by an individual in sample, we introduce latent variables
\begin{equation}
\label{eq:tlist}
\mathbf{T} =  \{T_{1},T_{2},\ldots,T_{n}\},
\end{equation}
for $T_k$ denoting the number of consecutive generations of selfing in the immediate ancestry of the $k^\emph{th}$ individual, and
\begin{equation}
\label{eq:ilist}
\mathbf{I} = \{I_{lk}\},
\end{equation}
for $I_{lk}$ indicating whether the lineages borne by the $k^\emph{th}$ individual at locus $l$ coalesce within the most recent $T_k$ generations.  Independent of other individuals, the number of consecutive generations of inbreeding
in the ancestry of the $k^\emph{th}$ individual is geometrically distributed:
\begin{equation}
\label{eq:T_k}
T_{k}  \sim  \mbox{Geometric}\left(s^{*}\right),
\end{equation}
with $T_{k}=0$ signifying that 
individual $k$ is the product of random outcrossing.  Irrespective of whether $0$, $1$, or $2$ of the genes at locus $l$ in individual $k$ are observed, $I_{lk}$ indicates whether the two genes at that locus in individual $k$ coalesce during the $T_k$ consecutive 
generations of inbreeding in its immediate ancestry:
\begin{equation*}
I_{lk} =  
\begin{cases}
0 & \mbox{if the two genes do not coalesce}\\
1 & \mbox{if the two genes coalesce.}
\end{cases}
\end{equation*}
Because the pair of lineages at any locus coalesce with probability $\frac{1}{2}$ in each generation of selfing,
\begin{equation}
\label{eq:Pr_I}
\Pr(I_{lk}=0) = \frac{1}{2^{T_{k}}}=1-\Pr(I_{lk}=1).
\end{equation}

\begin{figure*}
\begin{centering}
\includegraphics[width=0.68\textwidth]{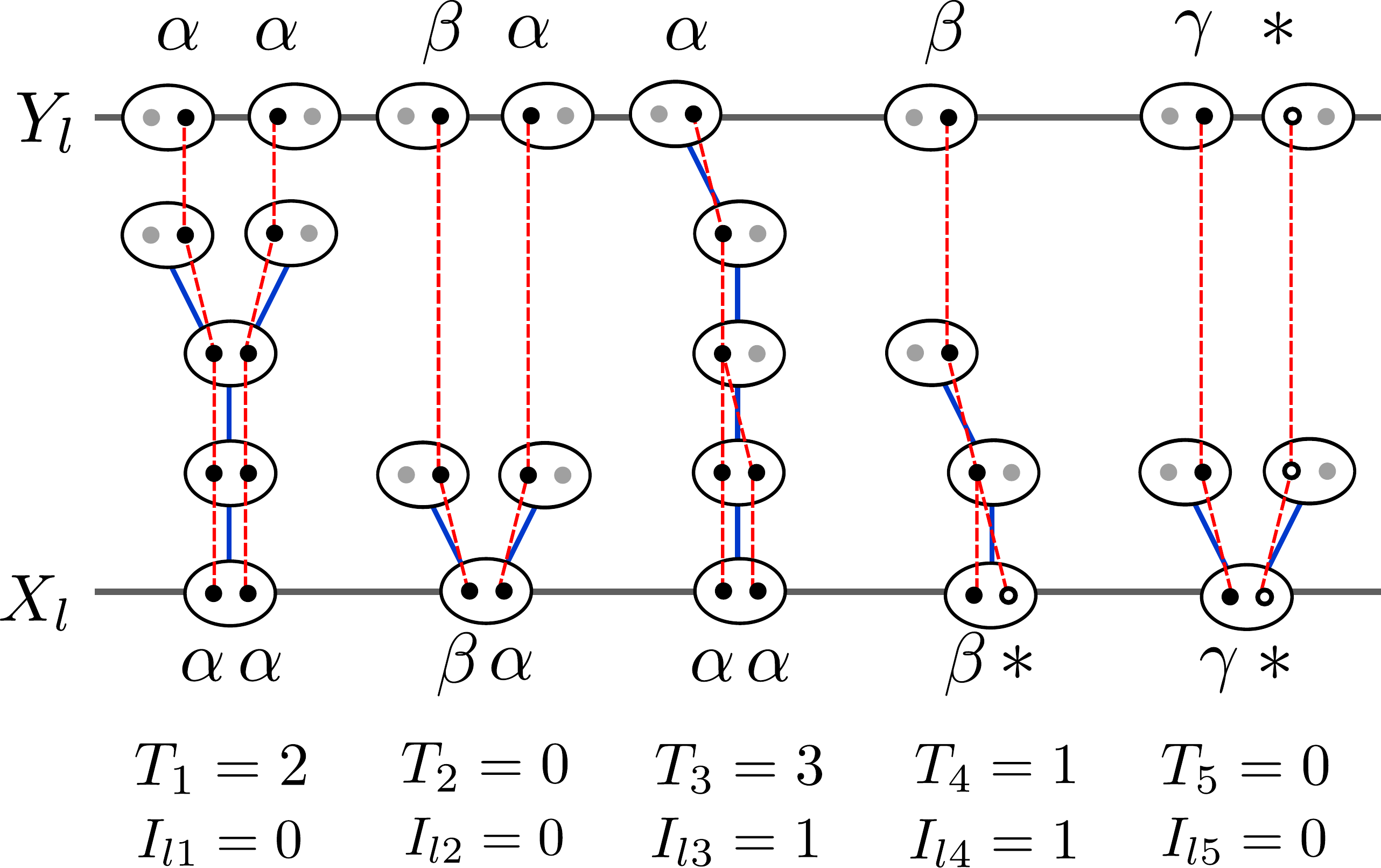}
\par\end{centering}

\protect\caption{\label{fig:Individuals-ancestries}  Following the
  history of the sample ($\mathbf{X}_l$) backwards in time until all
  ancestors of sampled genes reside in different individuals
  ($\mathbf{Y}_l$). Ovals represent
  individuals and dots represent genes.  Blue lines indicate the
  parents of individuals, while red lines represent the ancestry of
  genes. Filled dots represent sampled genes for which the allelic
  class is observed (Greek letters) and their ancestral lineages.
  Open dots represent genes in the sample with unobserved allelic
  class ($*$).  Grey dots represent other genes carried by ancestors
  of the sampled individuals.  The relationship between the observed
  sample $\mathbf{X}_{l}$ and the ancestral sample $\mathbf{Y}_{l}$ is
  determined by the intervening coalescence events
  $\mathbf{I}_{l}$. $\mathbf{T}$ indicates the number of consecutive
  generations of selfing for each sampled individual.}
\end{figure*}

Figure \ref{fig:Individuals-ancestries}
depicts the recent genealogical history at a locus $l$ in $5$ individuals.  Individuals $2$ and $5$ are products of random outcrossing ($T_{2}=T_{5}=0$), while the others derive from some positive number of consecutive generations of selfing in their immediate ancestry ($T_{1}=2, T_{3}=3, T_{4}=1$).  Both individuals $1$ and $3$ are homozygotes ($\alpha \alpha$), with the lineages of individual $3$ but not $1$ coalescing more recently than the most recent outcrossing event ($I_{l1}=0, I_{l3}=1$).  As individual $2$ is heterozygous ($\alpha \beta$), its lineages necessarily remain distinct since the most recent outcrossing event ($I_{l2}=0$).  One gene in each of individuals $4$ and $5$ are unobserved ($*$), with the unobserved lineage in individual $4$ but not $5$ coalescing more recently than the most recent outcrossing event ($I_{l4}=1, I_{l5}=0$).

In addition to the observed sample of diploid individuals, we consider the state of the sampled lineages at the most recent generation in which an outcrossing event has occurred in the ancestry of all $n$ individuals.  This point in the history of the sample occurs $\hat{T}$ generations into the past, for
\begin{equation*}
\hat{T} =  1+\max_{k}\, T_{k}.
\end{equation*}
In
Figure \ref{fig:Individuals-ancestries}, for example, $\hat{T}=4$,
reflecting the most recent outcrossing event in the ancestry of individual
$3$.  The ESF provides the probability of the allele frequency spectrum at this point.

We represent the ordered list of allelic states of the lineages at $\hat{T}$ generations into the past by
\begin{equation}
\label{eq:Y}
\mathbf{Y} = \left\{ \mathbf{Y}_{1},\mathbf{Y}_{2},\ldots,\mathbf{Y}_{L}\right\},
\end{equation}
for $\mathbf{Y}_{l}$ a list of ancestral genes
in the same order as their descendants in $\mathbf{X}_{l}$. 
Each gene in $\mathbf{Y}_{l}$ is the ancestor of either 1 or 2 genes at locus $l$ from
a particular individual in $\mathbf{X}_{l}$ \eqref{eq:X_l}, depending on whether the lineages held by that individual coalesce during the consecutive generations of inbreeding in its immediate ancestry.  We represent
the number of genes in $\mathbf{Y}_{l}$ by $m_{l}$ ($n \leq m_{l} \leq 2n$).   In Figure \ref{fig:Individuals-ancestries}, for example, $\mathbf{X}_{l}$ contains 10 genes in 5 individuals, but $\mathbf{Y}_{l}$
contains only $8$ genes, with $Y_{l1}$ the
ancestor of only the first allele of $\mathbf{X}_{l1}$ and $Y_{l5}$ the ancestor of both alleles of $\mathbf{X}_{l3}$.

We assume \eqref{eq:assumptions} that the initial phase of consecutive generations of selfing
is sufficiently short to ensure a negligible probability of mutation
in any lineage at any locus and a negligible probability of coalescence
between lineages held by distinct individuals more recently than $\hat{T}$.
Accordingly, the coalescence history $\mathbf{I}$ \eqref{eq:ilist} completely determines the correspondence
between genetic lineages in $\mathbf{X}$ \eqref{eq:X} and $\mathbf{Y}$ \eqref{eq:Y}.

\paragraph{Computing the likelihood:}  In principle, the likelihood of the observed data can be computed from the augmented
likelihood by summation:
\begin{equation}
\Pr(\mathbf{X}|\boldsymbol{\Theta}^{*},s^\ast) = \sum_{\mathbf{I}}\sum_{\mathbf{T}}\Pr(\mathbf{X},\mathbf{I},\mathbf{T}|\boldsymbol{\Theta}^{*},s^\ast),
\label{eq:likelihood}
\end{equation}
for
\begin{equation}
\label{eq:bigTheta}
\boldsymbol{\Theta}^{*} = \{\theta_1^\ast, \theta_2^\ast, \ldots, \theta_L^\ast\}
\end{equation}
the list of scaled, locus-specific mutation rates, $s^\ast$ the population-wide uniparental proportion for the reproductive system under consideration (\emph{e.g.}, \eqref{eq:pure-h} for the pure hermaphroditism model), and $\mathbf{T}$ \eqref{eq:tlist} and $\mathbf{I}$ \eqref{eq:ilist} the lists of latent variables representing the time since the most recent outcrossing event and whether the two lineages borne by a sampled individual coalesce during this period.  Here we follow a common abuse
of notation in using $\Pr(\mathbf{X})$ to denote $\Pr(\mathbf{X}=\mathbf{x})$
for random variable $\mathbf{X}$ and realized value $\mathbf{x}$.  Summation \eqref{eq:likelihood} is computationally expensive:  the number of consecutive generations of inbreeding in the immediate ancestry of an individual ($T_{k}$) has no upper limit \citep[compare][]{David2007} and the number of combinations of coalescence states ($I_{lk}$) across the $L$ loci and $n$ individuals increases exponentially ($2^{Ln}$) with the total number of assignments.  We perform Markov chain
Monte Carlo (MCMC) to avoid both these sums.

To calculate the augmented likelihood, we begin by applying Bayes
rule:
\begin{equation*}
\Pr(\mathbf{X},\mathbf{I},\mathbf{T}|\boldsymbol{\Theta}^{*},s^\ast) = \Pr(\mathbf{X},\mathbf{I}|\mathbf{T},\boldsymbol{\Theta}^{*},s^\ast)\Pr(\mathbf{T}|\boldsymbol{\Theta}^{*},s^\ast).
\end{equation*}
Because the times since the most recent outcrossing event $\mathbf{T}$ depend only on the uniparental proportion $s^\ast$, through \eqref{eq:T_k}, and not on the rates of mutation $\boldsymbol{\Theta}^{*}$, 
\begin{equation*}
\Pr(\mathbf{T}|\boldsymbol{\Theta}^{*},s^\ast)= \prod_{k=1}^{n} \Pr(T_{k}|s^\ast).
\end{equation*}

Even though our model assumes the absence of physical linkage among any of the loci, the genetic data $\mathbf{X}$ and coalescence
events $\mathbf{I}$ are not independent across loci because
they depend on the times since the most recent outcrossing event $\mathbf{T}$.  Given $\mathbf{T}$, however, the genetic data and coalescence events are independent across loci
\begin{equation*}
\label{eq:loci_independent_given_t}
\Pr(\mathbf{X},\mathbf{I}|\mathbf{T},\boldsymbol{\Theta}^{*},s^\ast)= \prod_{l=1}^{L} \Pr(\mathbf{X}_{l}, \mathbf{I}_{l}|\mathbf{T},\theta_{l}^{*}, s^\ast).
\end{equation*}
Further,
\begin{eqnarray*}
\Pr(\mathbf{X}_{l}, \mathbf{I}_{l}|\mathbf{T},\theta_{l}^\ast,s^\ast) & = & \Pr(\mathbf{X}_{l}|\mathbf{I}_{l},\mathbf{T},\theta_{l}^{*},s^\ast)\cdot\Pr(\mathbf{I}_{l}|\mathbf{T},\theta_{l}^{*},s^\ast)\\
 & = & \Pr(\mathbf{X}_{l}|\mathbf{I}_{l},\theta_{l}^{*}, s^\ast) \cdot \prod^n_{k=1} \Pr(I_{lk}|T_{k}).
\end{eqnarray*}
This expression reflects that the times to the most recent outcrossing event $\mathbf{T}$ affect the observed genotypes $\mathbf{X}_{l}$ only through the coalescence states $\mathbf{I}_{l}$ and that the coalescence states $\mathbf{I}_{l}$ depend only on the times to the most recent outcrossing event $\mathbf{T}$, through \eqref{eq:Pr_I}.

To compute $\Pr(\mathbf{X}_{l}|\mathbf{I}_{l},\theta_{l}^{*}, s^\ast)$, we incorporate latent variable $\mathbf{Y}_{l}$ \eqref{eq:Y}, describing the states of lineages at the most recent point at which all occur in distinct individuals (Figure \ref{fig:Individuals-ancestries}):
\begin{subequations}
\begin{eqnarray}
\Pr(\mathbf{X}_{l}|\mathbf{I}_{l},\theta_{l}^{*}, s^\ast) & = & \sum_{\mathbf{Y}_{l}}\Pr(\mathbf{X}_{l},\mathbf{Y}_{l}|\mathbf{I}_{l},\theta_{l}^{*}, s^\ast)\nonumber \\
& = & \sum_{\mathbf{Y}_{l}}\Pr(\mathbf{X}_{l} | \mathbf{Y}_{l}, \mathbf{I}_{l},\theta_{l}^{*}, s^\ast)\Pr(\mathbf{Y}_{l} | \mathbf{I}_{l},\theta_{l}^{*}, s^\ast)\nonumber \\
 & = & \sum_{\mathbf{Y}_{l}}\Pr(\mathbf{X}_{l}|\mathbf{Y}_{l},\mathbf{I}_{l})\cdot\Pr(\mathbf{Y}_{l}|\mathbf{I}_{l},\theta_{l}^{*}),\label{eq:bayes2}
\end{eqnarray}
reflecting that the coalescence states $\mathbf{I}_{l}$ establish the correspondence between the spectrum of genotypes in $\mathbf{X}_{l}$ and the spectrum of alleles in $\mathbf{Y}_{l}$ and that the distribution of $\mathbf{Y}_{l}$, given by the ESF, depends on the uniparental proportion $s^\ast$ only through the scaled mutation rate $\theta_l^\ast$ \eqref{eq:theta-star}.

Given the sampled genotypes $\mathbf{X}_{l}$ and coalescence states $\mathbf{I}_{l}$, at most one ordered list of alleles $\mathbf{Y}_{l}$ produces positive $\Pr(\mathbf{X}_{l}|\mathbf{Y}_{l},\mathbf{I}_{l})$ in \eqref{eq:bayes2}.  Coalescence of the lineages at locus $l$ in any heterozygous individual (\emph{e.g.}, $X_{lk}=(\beta,\alpha)$ with $I_{lk}=1$ in Figure \ref{fig:Individuals-ancestries}) implies
\begin{equation*}
\Pr(\mathbf{X}_{l}|\mathbf{Y}_{l},\mathbf{I}_{l})=0
\end{equation*}
for all $\mathbf{Y}_{l}$.  Any non-zero $\Pr(\mathbf{X}_{l}|\mathbf{Y}_{l},\mathbf{I}_{l})$ precludes coalescence in any heterozygous individual and $\mathbf{Y}_{l}$ must specify the observed alleles of $\mathbf{X}_{l}$ in the order of observation, with either 1 ($I_{lk}=1$) or 2 ($I_{lk}=0$) instances of the allele for any homozygous individual (\emph{e.g.}, $X_{lk}=(\alpha,\alpha)$).
For all cases with non-zero $\Pr(\mathbf{X}_{l}|\mathbf{Y}_{l},\mathbf{I}_{l})$,
\begin{equation*}
\Pr(\mathbf{X}_{l}|\mathbf{Y}_{l},\mathbf{I}_{l})=1.
\end{equation*}
Accordingly, expression (\ref{eq:bayes2}) reduces to
\begin{equation}
\label{eq:bayes2short}
\Pr(\mathbf{X}_{l}|\mathbf{I}_{l},\theta_{l}^{*}, s^\ast) =
\sum_{\mathbf{Y}_{l}: \Pr(\mathbf{X}_{l}|\mathbf{Y}_{l},\mathbf{I}_{l}) \neq 0} \Pr(\mathbf{Y}_{l}|\mathbf{I}_{l},\theta_{l}^{*}),
\end{equation}
\end{subequations}
a sum with either 0 or 1 terms.  
Because all genes in $\mathbf{Y}_{l}$ reside in distinct individuals, we obtain $\Pr(\mathbf{Y}_{l}|\mathbf{I}_{l},\theta_{l}^{*})$ from the Ewens Sampling Formula for a sample, of size
\begin{equation*}
m_{l}=2n-\sum_{k=1}^{n}I_{lk},
\end{equation*}
ordered in the sequence in which the genes are observed.

To determine $\Pr_{}(\mathbf{Y}_{l}|\mathbf{I}_{l},\theta_{l}^{*})$
in \eqref{eq:bayes2short}, we use a fundamental property of the ESF \citep{Ewens1972,Karlin1972}:  the probability that the next-sampled ($i^{th}$) gene represents a novel allele corresponds to
\begin{subequations}
\label{nextgenei}
\begin{equation}
\label{novelpi}
\pi_{i} = \frac{\theta^{*}}{i-1+\theta^{*}},
\end{equation}
for $\theta^{*}$ defined in (\ref{eq:theta-star}), and the probability
that it represents an additional copy of already-observed allele $j$ is
\begin{equation}
(1-\pi_{i})\frac{i_{j}}{i-1},
\end{equation}
\end{subequations}
for $i_{j}$ the number of replicates of allele $j$ in the sample at size $(i-1)$
($\sum_{j}i_{j}=i-1$).  Appendix \ref{nextsampledgene} presents a first-principles derivation of \eqref{novelpi}.  Expressions \eqref{nextgenei} imply that for $\mathbf{Y}_l$ the list of alleles at locus $l$ in order of observance,
\begin{equation}
\label{proby}
\Pr(\mathbf{Y}_{l}|\mathbf{I}_{l},\theta_l^{*})= \frac{(\theta_l^{*})^{K_l}\prod_{j=1}^{K_l}(m_{lj}-1)!}{\prod_{i=1}^{m_{l}}(i-1+\theta_l^{*})},
\end{equation}
in which $K_l$ denotes the total number of distinct allelic classes, $m_{lj}$ the number of replicates of the $j^{th}$ allele in the sample, and $m_{l}=\sum_{j}m_{lj}$ the number of lineages remaining at time $\hat{T}$ (Figure \ref{fig:Individuals-ancestries}).

\paragraph{Missing data:}  Our method allows the allelic class of one or both genes at each locus to be missing.  In Figure \ref{fig:Individuals-ancestries}, for example, the genotype
of individual\textbf{ }$4$ is $\mathbf{X}_{l4}=(\beta,*)$, indicating
that the allelic class of the first gene is observed to be $\beta$, but that of the second gene
is unknown.

A missing allelic specification in the sample of genotypes $\mathbf{X}_{l}$ leads to a missing specification for
the corresponding gene in $\mathbf{Y}_{l}$ unless the genetic lineage
coalesces, in the interval between $\mathbf{X}_{l}$ and $\mathbf{Y}_{l}$, with a lineage ancestral to a gene for which the allelic type was observed.
Figure \ref{fig:Individuals-ancestries}
illustrates such a coalescence event in the case of individual $4$. In contrast, the lineages ancestral to the genes carried by individual $5$ fail to coalescence more recently than their separation into distinct individuals, giving rise to a missing specification in $\mathbf{Y}_{l}$.

The probability of $\mathbf{Y}_{l}$ can be computed by simply
summing over all possible values for each missing specification.  Equivalently, those elements may simply be dropped from $\mathbf{Y}_{l}$ before computing the probability via the ESF, the procedure implemented in our method.

\section*{Bayesian inference framework}

\subsection*{\label{sub:Prior-on-mutation}Prior on mutation rates}

\citet{Ewens1972} showed for the panmictic case that the number of distinct allelic classes observed at a locus (\emph{e.g.}, $K_l$ in \eqref{proby}) provides a sufficient statistic for the estimation of the scaled mutation rate. Because each locus $l$ provides relatively little information about the scaled mutation rate $\theta_{l}^{*}$ \eqref{eq:theta-star}, we assume that mutation rates across loci cluster in a finite number of groups.  However, we do not know \emph{a priori} the group assignment of loci or even the number of distinct rate classes among the observed loci. We make use of the Dirichlet process prior to estimate simultaneously the number of groups, the value of $\theta^{*}$
for each group, and the assignment of loci to groups.

The Dirichlet process comprises a base distribution, which here represents the distribution of the scaled mutation rate $\theta^\ast$ across groups, and a concentration parameter $\alpha$, which controls the probability that each successive locus forms a new group.  We assign $0.1$ to $\alpha$ of the Dirichlet process, and place a gamma distribution ($\Gamma(\alpha=0.25,\beta=2)$) on the mean scaled mutation rate for each group.  As this prior has a high variance
relative to the mean (0.5), it is relatively uninformative
about $\theta^{*}$.

\subsection*{Model-specific parameters}\label{param}

Derivations presented in the preceding section indicate that the probability of a sample of diploid genotypes under the infinite alleles model depends on only the uniparental proportion $s^\ast$ and the scaled mutation rates $\boldsymbol{\Theta}^\ast$ \eqref{eq:bigTheta} across loci.  These composite parameters are determined by the set of basic demographic parameters $\mathbf{\Psi}$ associated with each model of reproduction under consideration.  As the genotypic data provide equal support to any combination of basic parameters that implies the same values of $s^\ast$ and $\boldsymbol{\Theta}^\ast$, the full set of basic parameters for any model are in general non-identifiable using the observed genotype frequency spectrum alone.

Even so, our MCMC implementation updates the basic parameters directly, with likelihoods determined from the implied values of $s^\ast$ and $\boldsymbol{\Theta}^\ast$.  This feature facilitates the incorporation of information in addition to the genotypic data that can contribute to the estimation of the basic parameters under a particular model or assessment of alternative models.
We have
\begin{align}
\label{likeparameters}
\Pr(\mathbf{X},\mathbf{\Theta}^{*},\mathbf{\Psi}) & = \Pr(\mathbf{X}|\mathbf{\Theta}^{*},\mathbf{\Psi})\cdot\Pr(\mathbf{\Theta}^{*})\cdot\Pr(\mathbf{\Psi}) \nonumber \\
 & = \Pr(\mathbf{X}|\mathbf{\Theta}^{*},s^{*}(\mathbf{\Psi}))\cdot\Pr(\mathbf{\Theta}^{*})\cdot\Pr(\mathbf{\Psi}),
\end{align}
for $\mathbf{X}$ the genotypic data and $s^{*}(\mathbf{\Psi})$ the uniparental proportion determined by $\mathbf{\Psi}$ for the model under consideration. To determine the marginal distribution of $\theta_{l}$ \eqref{rates}
for each locus $l$, we use \eqref{eq:theta-star}, incorporating the distributions of $s^{*}(\mathbf{\Psi})$ and $S(\mathbf{\Psi})$, the scaling factor defined in \eqref{rates}:
\begin{equation*}
\theta_l =  \frac{\theta^\ast_l}{S(1-s^\ast/2)}.
\end{equation*}

For the pure hermaphroditism model \eqref{eq:pure-h}, $\mathbf{\Psi} = \{\tilde s,\tau\}$, where $\tilde s$ is the proportion of conceptions through selfing, and $\tau$ is the relative viability of uniparental offspring.  We propose uniform priors for $\tilde s$ and $\tau$:
\begin{equation}
\label{parampureh}
\begin{split}
\tilde s & \sim \text{Uniform}(0,1) \\
\tau     & \sim \text{Uniform}(0,1).
\end{split}
\end{equation}
For the androdioecy model \eqref{eq:androdioecy}, we propose uniform priors for each basic parameter in $\mathbf{\Psi}=\{\tilde s, \tau, p_{m}\}$:
\begin{equation}
\label{paramandro}
\begin{split}
\tilde s & \sim \text{Uniform}(0,1) \\
\tau     & \sim \text{Uniform}(0,1)\\
    p_{m}  & \sim \text{Uniform}(0,1).
\end{split}
\end{equation}
For the gynodioecy model \eqref{eq:gynodioecy}, $\mathbf{\Psi}=\{a, \tau, p_{f}, \sigma \}$, including $a$ the proportion of egg cells produced by hermaphrodites fertilized by selfing, $p_{f}$ \eqref{definerhof} the proportion of females (male-steriles) among reproductives, and $\sigma$ the fertility of females relative to hermaphrodites.  We propose the uniform priors
\begin{equation}
\label{paramgyno}
\begin{split}
a & \sim \text{Uniform}(0,1) \\
\tau     & \sim \text{Uniform}(0,1)\\
    p_{f} & \sim \text{Uniform}(0,1)\\
    1/\sigma & \sim \text{Uniform}(0,1).
\end{split}
\end{equation}

\section*{Assessment of accuracy and coverage using simulated data}

We developed a forward-in-time simulator (\url{https://github.com/skumagai/selfingsim}) that tracks multiple neutral loci with locus-specific scaled mutation rates ($\boldsymbol{\Theta}$) in a population comprising $N$ reproducing hermaphrodites of which a proportion $s^\ast$ are of uniparental origin.  We used this simulator to generate data under two sampling regimes:  large ($L=32$ loci in each of $n=70$ diploid individuals) and small ($L=6$ loci in each of $n=10$ diploid individuals).  We applied our Bayesian method and \texttt{RMES} \citep{David2007} to simulated data sets.  A description of the procedures used to assess the accuracy and coverage properties of the three methods is included in the Supplementary Online Material.

In addition, we determine the uniparental proportion ($s^\ast$) inferred from the departure from Hardy-Weinberg expectation \citep[$F_{IS}$,][]{Wright69} alone.  Our $F_{IS}$-based estimate entails setting the observed value of $F_{IS}$ equal to its classical expectation $s^\ast/(2-s^\ast)$ \citep{Wright1921, Haldane1924} and solving for $s^\ast$:
\begin{equation}
\label{fis}
\widehat{s^\ast} = \frac{2\widehat{F_{IS}}}{1+\widehat{F_{IS}}}.
\end{equation}
In accommodating multiple loci, this estimate incorporates a multilocus estimate for $\widehat{F_{IS}}$ (Appendix \ref{appfis}) but, unlike those generated by our Bayesian method and \texttt{RMES}, does not use identity disequilibrium across loci within individuals to infer the number of generations since the most recent outcross event in their ancestry.  As our primary purpose in examining the $F_{IS}$-based estimate \eqref{fis} is to provide a baseline for the results of those likelihood-based methods, we have not attempted to develop an index of error or uncertainty for it.

\subsection*{Accuracy}

To assess relative accuracy of estimates of the uniparental proportion $s^{*}$, we determine the bias and root-mean-squared error of the three methods by averaging over $10^4$ data sets ($10^2$ independent samples from each of $10^2$ independent simulations for each assigned $s^{*}$).  In contrast with the point estimates of $s^\ast$ produced by \texttt{RMES}, our Bayesian method generates a posterior distribution.  To facilitate comparison, we reduce our estimate to a single value, 
the median of the posterior distribution of $s^\ast$, with the caveat that the mode and mean may show different qualitative behavior (see Supplementary Online Material).

\begin{figure}
\begin{centering}
\includegraphics[width=0.49\textwidth]{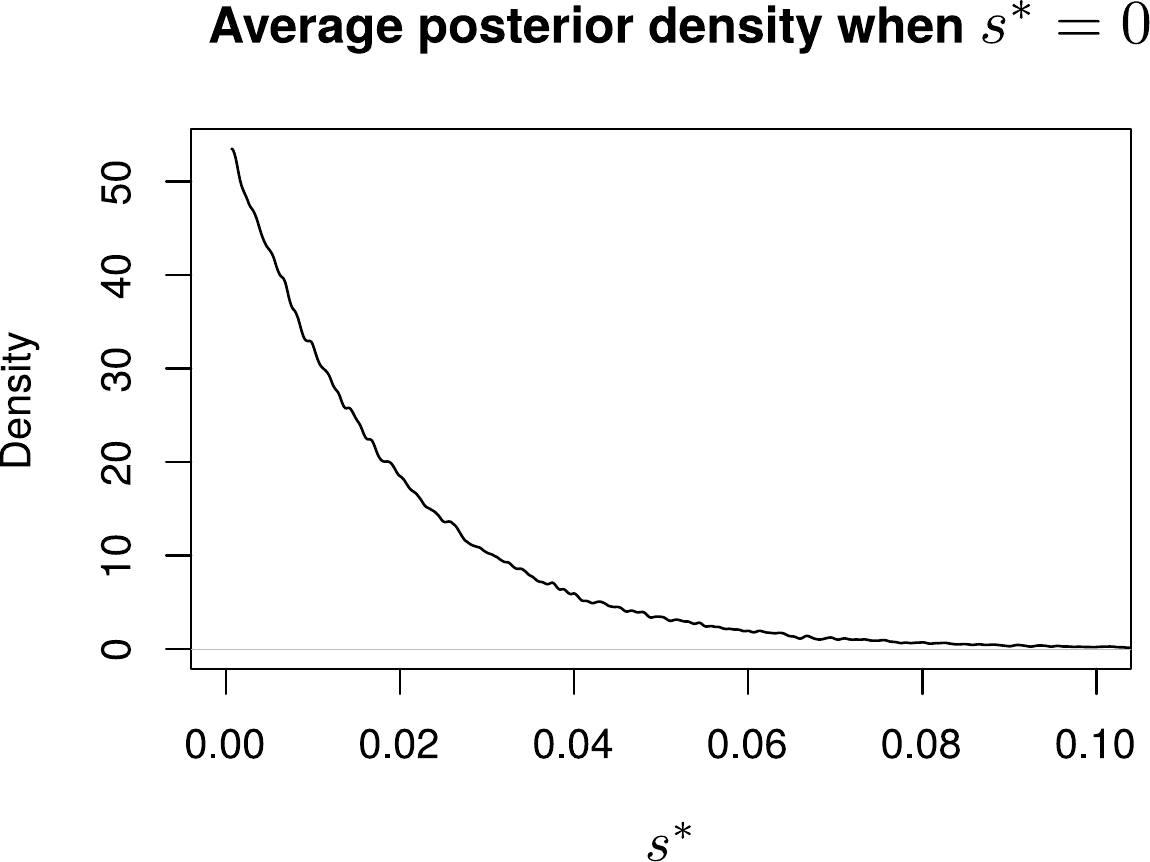}
\protect\caption{\label{fig:Average-posterior-density-s0}Average posterior density of the uniparental proportion ($s^{*}$) inferred from simulated data generated under the large sample regime ($n=70$, $L=32$) with a true value of $s^{*}=0$. The average
was taken across posterior densities for 100 data sets.}

\par\end{centering}

\end{figure}

Figure \ref{fig:error-posterior-median} indicates that both \texttt{RMES} and our method show positive bias upon application to data sets for which the true uniparental proportion $s^{*}$ is close to zero and negative bias for $s^{*}$ 
\begin{figure}[h!]
\begin{centering}
\includegraphics[width=\textwidth]{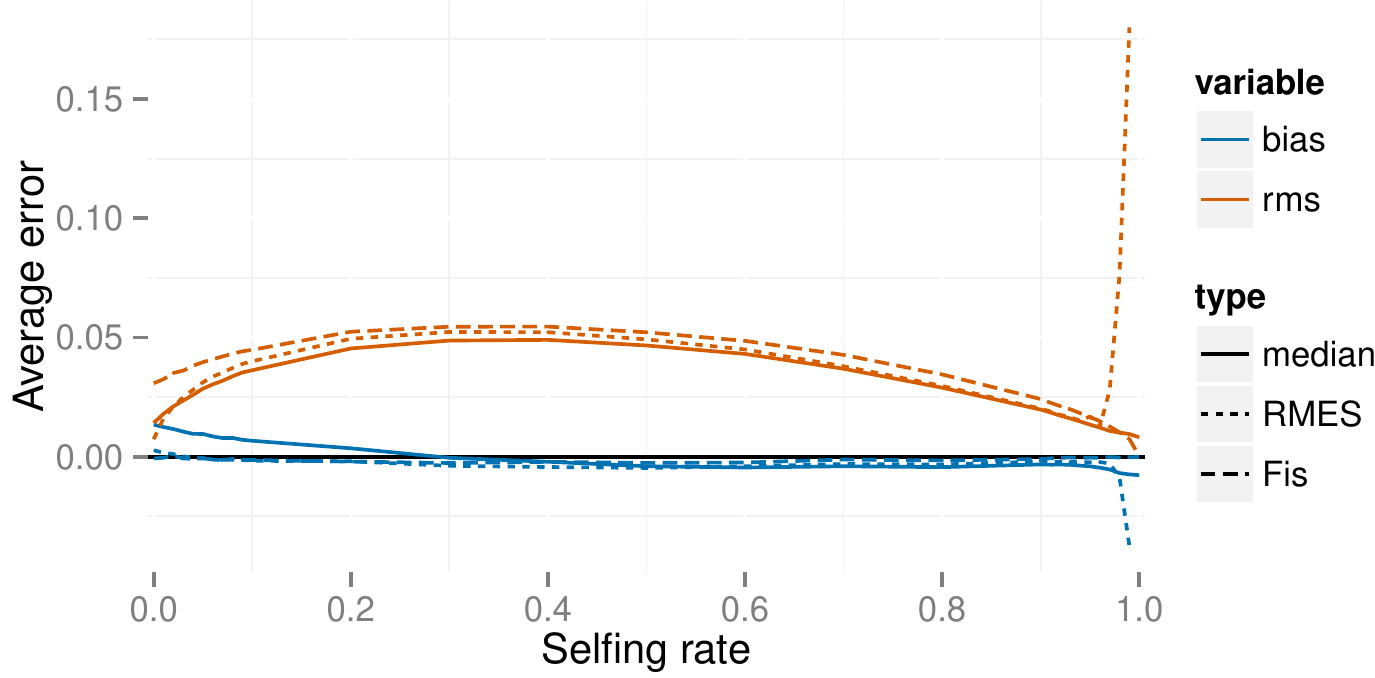}
\par\end{centering}
\protect\caption{\label{fig:error-posterior-median}Errors for the full likelihood
(posterior median), \texttt{RMES}, and $F_{IS}$-based \eqref{fis} methods for a large simulated sample ($n=70$ individuals, $L=32$ loci). In the legend, rms indicates the
root-mean-squared error and bias the average deviation.
Averages are taken across simulated data sets at each true value of
$s^{*}$.}
\end{figure}
close to unity. This trend reflects that both methods yield estimates of $s^{*}$ constrained to lie between $0$ and $1$. In contrast, the $F_{IS}$-based estimate \eqref{fis} underestimates $s^{*}$ throughout the range, even near $s^{*}=0$ ($\widehat{F_{IS}}$ is not constrained to be positive).  Our method
has a bias near $0$ that is substantially larger than the bias of
\texttt{RMES}, and an error that is slightly larger. A major contributor to this trend is that our Bayesian estimate is represented by only the median of the posterior distribution of the uniparental proportion 
$s^\ast$.
Figure \ref{fig:Average-posterior-density-s0} indicates that for data sets generated under a true value of $s^{*}$ of $0$ (full random outcrossing), 
the posterior distribution for $s^{*}$ has greater mass near $0$.  Further, as the posterior mode does not display large bias near $0$ (Figure \ref{fig:error-posteriors}), we conclude
that the bias shown by the median (Figure \ref{fig:error-posterior-median}) merely represents uncertainty in the posterior distribution for $s^{*}$ and not any preference for incorrect values.  We note that our method assumes that the data are derived from a population reproducing through a mixture of self-fertilization and random outcrossing.  Assessment of a model of complete random mating ($s^\ast=0$) against the present model ($s^\ast>0$) might be conducted through the Bayes factor.

\begin{figure}[h!]
\begin{centering}
\includegraphics[width=0.48\textwidth]{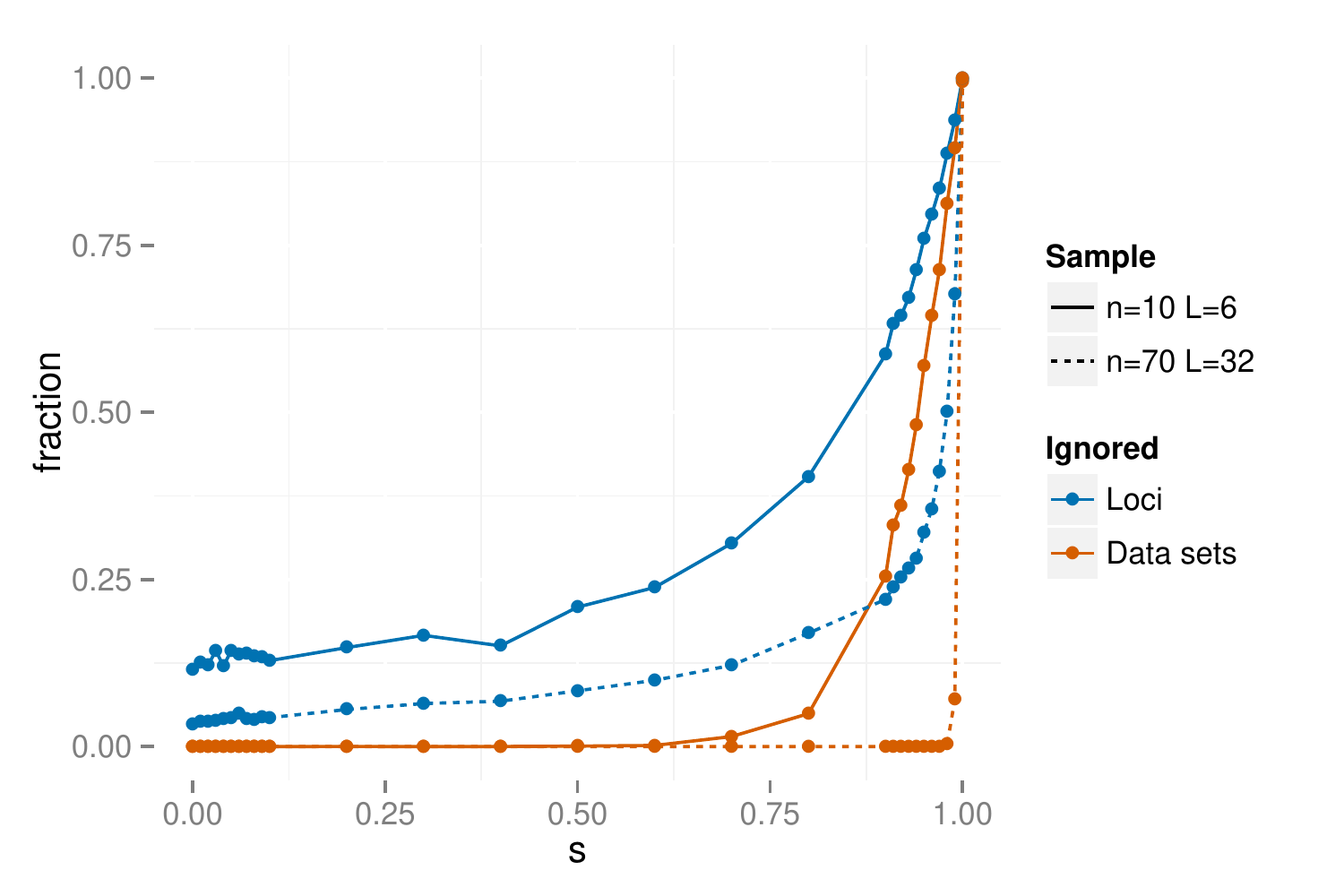}
\par\end{centering}

\protect\caption{\label{fig:Loci-and-data-unusable-by-RMES}Fraction of loci and data
sets that are ignored by \texttt{RMES}.}
\end{figure}

Except in cases in which the true $s^{*}$ is very close to
$0$,  the error for \texttt{RMES} exceeds the error for our method
under both sampling regimes (Figure \ref{fig:error-posterior-median}).
\texttt{RMES} differs from the other two methods in the steep rise in both bias and rms error for high values of $s^{*}$, with the change point occurring at lower values of the uniparental proportion $s^\ast$ for the small sampling regime ($n=10$, $L=6$).  A likely contributing factor to the increased error shown by  \texttt{RMES} under high values of $s^{*}$ is its default assumption that the number of generations in the ancestry of any individual does not exceed $20$.  Violations of this assumption arise more often under high values of $s^{*}$, possibly promoting underestimation of the uniparental proportion.  Further, \texttt{RMES}
discards data at loci at which no heterozygotes are observed, and terminates analysis altogether if the number of loci drops below $2$.  \texttt{RMES} treats all loci with zero heterozygosity \eqref{apphet} as uninformative, even if multiple alleles are observed.  In contrast, our full likelihood method uses data from all loci, with polymorphic loci in the absence of heterozygotes providing strong evidence of high rates of selfing (rather than low rates of mutation).  Under the large sampling
regime ($n=70$, $L=32$), \texttt{RMES} discards on average 50\% of the loci for true $s^{*}$ values exceeding $0.94$, with less than $10\%$ of data sets unanalyzable (fewer than 2 informative loci) even at $s^{*}=0.99$
(Figure \ref{fig:Loci-and-data-unusable-by-RMES}). 
 Under the $n=10$, $L=6$ regime, \texttt{RMES} discards on average 50\% of loci for true $s^{*}$ values exceeding $0.85$, with about 50\% of data sets  unanalyzable under $s^{*}\ge0.94$.

The error
for the $F_{IS}$-based estimate \eqref{fis} also exceeds the error for our method. It is largest near $s^{*}=0$ and
vanishes as $s^{*}$ approaches $1$, a pattern distinct from \texttt{RMES} (Figure \ref{fig:error-posterior-median}).

\subsection*{Coverage}

We determine the fraction of data sets for which the confidence interval (CI) generated by \texttt{RMES} and the Bayesian credible interval (BCI) generated by
our method contains the true value of the uniparental proportion $s^{*}$. This measure of coverage
is a frequentist notion, as it treats each true value of $s^{*}$
separately. A 95\% CI should contain the truth 95\% of the time for
each specific value of $s^{*}$. However, a 95\% BCI is not expected to have
95\% coverage at each value of $s^{*}$, but rather 95\%
coverage averaged over values of $s^{*}$ sampled from the prior.  Of the various ways to determine a BCI for a given posterior
distribution, we choose to report the highest posterior density BCI (rather than the central BCI, for example).

\begin{figure*}[h!]
\begin{centering}
\includegraphics[width=0.48\textwidth]{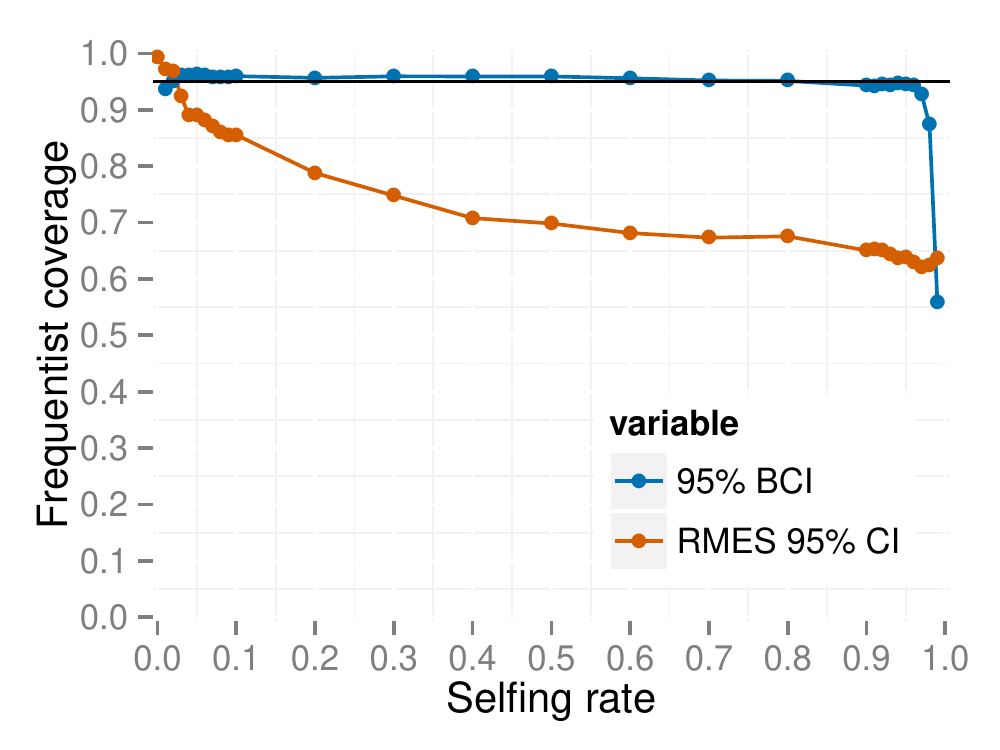}
\par\end{centering}
\protect\caption{\label{fig:Comparison-RMES-CI}Frequentist coverage at each level
of $s^{*}$ for 95\% intervals from \texttt{RMES} and the method based on the
full likelihood under the large sampling regime ($n=70, L=32$). \texttt{RMES} intervals are 95\% confidence intervals computed
via profile likelihood. Full likelihood intervals are 95\% highest posterior density Bayesian
credible intervals.}
\end{figure*}

Figure \ref{fig:Comparison-RMES-CI} indicates that coverage of the 95\% CIs produced by \texttt{RMES} are consistently lower than 95\% across all true $s^{*}$ values under the large sampling regime ($n=70$ $L=32$).  Coverage appears to 
decline as $s^\ast$ increases, dropping from $86\%$ for $s^\ast=0.1$ to 64\% for $s^\ast=0.99$. In contrast, the 95\% BCIs have slightly greater than 95\% frequentist
coverage for each value of $s^{*}$, except for $s^{*}$ values very close to the extremes ($0$ and $1$).  Under very high rates of inbreeding ($s^{*} \approx 1$), an assumption  \eqref{eq:assumptions} of our underlying model (random outcrossing occurs on a time scale much shorter than the time scales of mutation and coalescence) is likely violated.  We observed similar behavior under nominal coverage levels ranging from $0.5$ to $0.99$ (Supplementary Material).

\subsection*{Number of consecutive generations of selfing}

In order to check the accuracy of our reconstructed generations of
selfing, we examine the posterior distributions of selfing times $\{T_{k}\}$
for $s^{*}=0.5$ under the large sampling regime ($n=70,L=32$). We average
\begin{figure}[h!]
\centering{}
\includegraphics[width=0.48\textwidth]{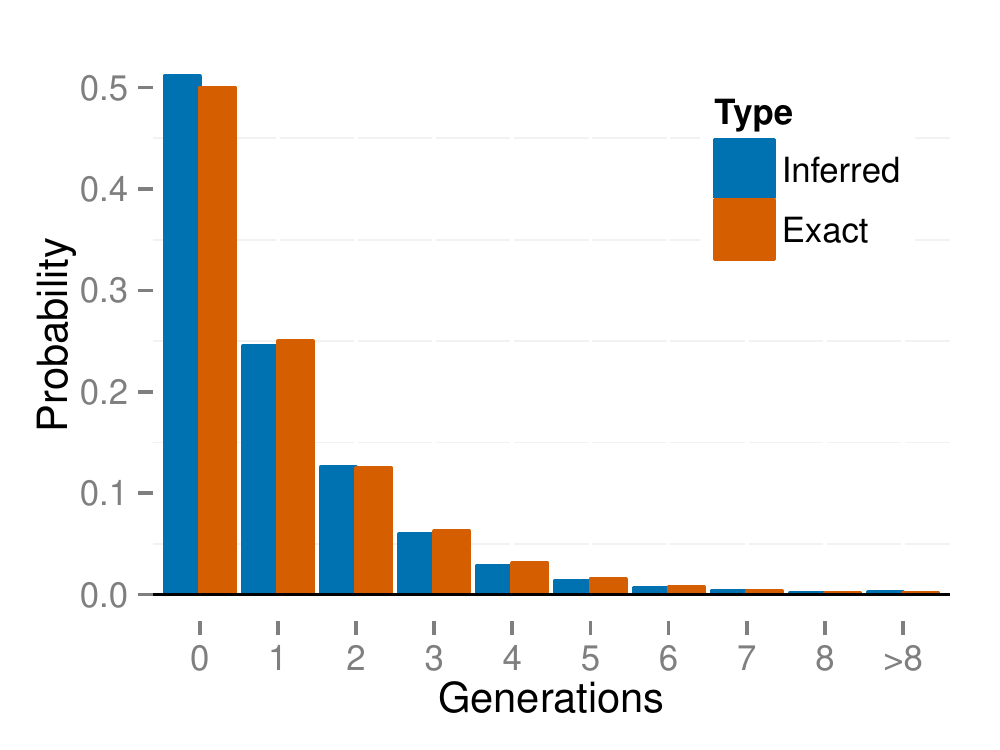}
\protect\caption{\label{fig:reconstructed-selfing-times-1}Exact
distribution of selfing
times under $s^\ast=0.5$ compared to the posterior distribution averaged across individuals
and across data sets.}
\end{figure}
posterior distributions for selfing times across 100 simulated data
sets, and across individuals $k=1\ldots70$ within each simulated
data set. We then compare these averages based on the simulated data
with the exact distribution of selfing times across individuals (Figure
\ref{fig:reconstructed-selfing-times-1}). 
 The pooled posterior distribution closely matches the exact distribution.
This simple check suggests that our method correctly infers the true
posterior distribution of selfing times for each sampled individual.

\section*{Analysis of microsatellite data from natural populations}

\subsection*{Androdioecious vertebrate}

Our analysis of data from the androdioecious killifish \emph{Kryptolebias marmoratus} \citep{Mackiewicz2006b, TatarenkovEarleyTaylorAvise2012} incorporates genotypes from 32 microsatellite loci as well as information on the observed fraction of males.  Our method simultaneously estimates the proportion of males in the population ($p_{m}$) together with rates of locus-specific mutation ($\theta^\ast$) and the uniparental proportion ($s_A$).  We apply the method to two populations, which show highly divergent rates of inbreeding. 

\paragraph{Parameter estimation:}  Our androdioecy model \eqref{parampureh} comprises 3 basic parameters, including the fraction of males among reproductives ($p_m$) and the relative viability of uniparental offspring ($\tau$).  Our analysis incorporates the observation of $n_m$ males among $n_\emph{total}$ zygotes directly into the likelihood expression:
\begin{equation*}
\Pr(\mathbf{X},\mathbf{I},\mathbf{T},n_m|s^\ast,\mathbf{\Theta}^{*}, p_{m}, n_\emph{total}) = \Pr(\mathbf{X}\mathbf{,}\mathbf{I},\mathbf{T}|s^{*},\mathbf{\Theta}^{*})\cdot\Pr(n_m|p_{m}, n_\emph{total}),
\end{equation*}
in which
\begin{equation}
\label{binmale}
n_m \sim \text{Binomial}(n_\emph{total}, p_{m}),
\end{equation}
reflecting that $s^{*}$ and $\mathbf{\Theta}^{*}$ are sufficient to account for $\mathbf{X}$, $\mathbf{I}$, and $\mathbf{T}$, and also independent of $n_{m}$, $n_\emph{total}$, and $p_{m}$.

In the absence of direct information regarding the existence or intensity of inbreeding depression, we impose the constraint $\tau=1$ to permit estimation of the uniparental proportion $s_A$ under a uniform prior:
\begin{equation*}
s^\ast  \sim \text{Uniform}(0,1).
\end{equation*}

\paragraph{Low outcrossing rate:} We applied our method to the BP data set described by \citet{TatarenkovEarleyTaylorAvise2012}.  This data set comprises a total of $70$ individuals, collected in $2007$, $2010$, and $2011$ from the Big Pine location on the Florida Keys.

\citet{TatarenkovEarleyTaylorAvise2012} report $21$ males among the $201$ individuals collected from various locations in the Florida Keys during this period, consistent with other estimates of about $1\%$ \citep[\emph{e.g.},][]{TurnerDavisTaylor1992}.  Based on the long-term experience of the Tatarenkov--Avise laboratory with this species, we assumed observation of $n_m=20$ males out of $n_\emph{total}=2000$ individuals in \eqref{binmale}.  We estimate that the fraction of males in the population ($p_{m}$) has a posterior median of $0.01$ with
a 95\% Bayesian
Credible Interval (BCI) of $(0.0062,0.015)$.

Our estimates of mutation rates ($\theta^\ast$) indicate substantial
variation among loci, with the median ranging over an order of magnitude (ca.\ $0.5$--$5.0$) (Figure \ref{fig:Kmar-BP-thetas-1}, Supplementary Material).  The distribution of mutation rates across loci appears to be multimodal, with many loci having a relatively low rate and some having larger rates.

\begin{figure*}[h!]
\begin{centering}
\includegraphics[width=0.48\textwidth]{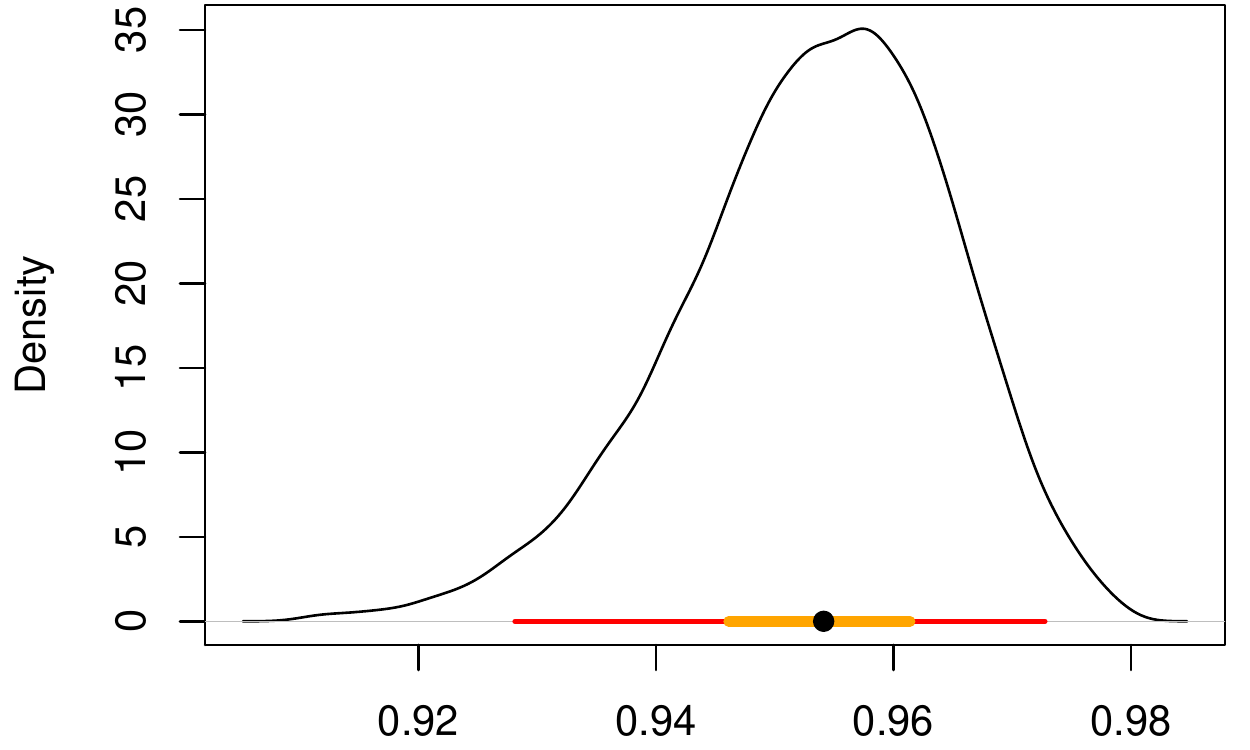}
\par\end{centering}
\protect\caption{\label{fig:Posterior-distributions-BP}Posterior distribution of the uniparental proportion $s_A$ for the BP population.   The median is indicated by a black dot, with a red bar for the 95\% BCI and an orange bar for the 50\% BCI.}
\end{figure*}

Figure \ref{fig:Posterior-distributions-BP} shows the posterior distribution of uniparental proportion
$s_A$, with a median of $0.95$ and a $95\%$ BCI of $(0.93,0.97)$.
This estimate is somewhat lower than $F_{IS}$-based estimate \eqref{fis} of $0.97$, and slightly higher than the \texttt{RMES} estimate of $0.94$, which has a 95\% Confidence Interval (CI) of $(0.91,0.96)$. We note that \texttt{RMES} discarded from the analysis 9 loci (out of 32) which showed no heterozygosity, even though 7 of the 9 were polymorphic in the sample.

\begin{figure*}
\includegraphics[width=0.96\textwidth]{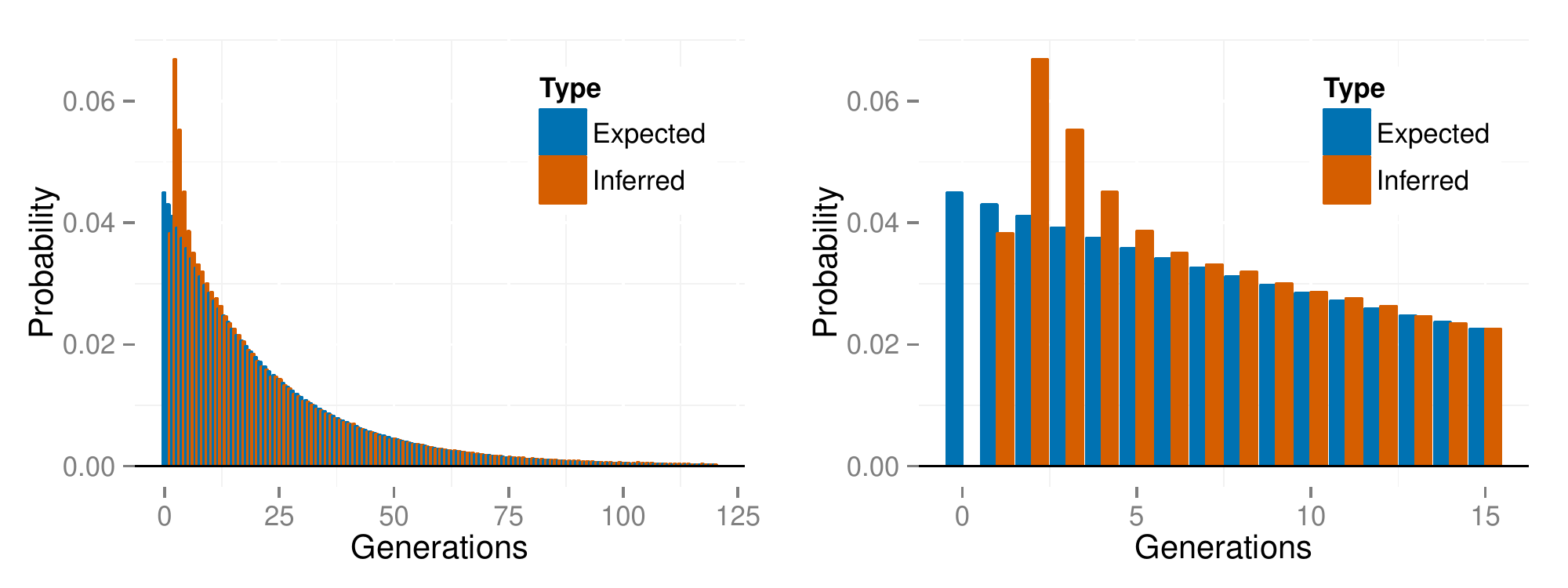}

\protect\caption{\label{fig:Kmar-BP-t-dist-2-1}Empirical distribution of number of generations since the most recent outcross event ($T$) across individuals for the \emph{K. marmoratus} (BP population), averaged across posterior samples. The right panel
is constructed by zooming in on the panel on the left.  ``Expected'' probabilities represent the proportion of individuals with the indicated number of selfing generations expected under the estimated uniparental proportion $s_A$.  ``Inferred'' probabilities represent proportions inferred across individuals in the sample.  The first inferred bar
with positive probability corresponds to $T=1$.}
\end{figure*}

Our method estimates the latent variables $\{T_1, T_2, \ldots, T_n \}$  \eqref{eq:tlist}, representing the number of generations since the most recent outcross event in the ancestry of each individual (Figure \ref{fig:Kmar-BP-t-dist-1}).  Figure \ref{fig:Kmar-BP-t-dist-2-1}
shows the empirical distribution of the time since outcrossing across
individuals, averaged over posterior uncertainty, indicating a complete absence of biparental individuals (0 generations of selfing). 
Because we expect that a sample of size 70 would include at least some biparental individuals under the inferred uniparental proportion
($s_A \approx 0.95$), this finding suggests that any biparental individuals in the sample show lower heterozygosity than expected from the observed level of genetic variation.
This deficiency suggests that an extended model that accommodates biparental inbreeding or population subdivision may account for the data better than the present model, which allows only selfing and random outcrossing.

\paragraph{Higher outcrossing rate:} We apply the three methods to the sample collected in $2005$ from Twin Cays, Belize \citep[TC05:][]{Mackiewicz2006b}.  This data set departs sharply from that of the BP population, showing considerably higher incidence of males and levels of polymorphism and heterozygosity. 

We incorporate the observation of $19$ males among the $112$ individuals collected from Belize in $2005$ \citep{Mackiewicz2006b} into the likelihood (see \eqref{binmale}).  Our estimate of the fraction of males in the population ($p_{m}$) has a posterior median of $0.17$ with
a 95\% BCI of $(0.11,0.25)$.

Figure \ref{fig:Kmar-PG-thetas-1-1} (Supplementary Material) indicates that the posterior medians of the locus-specific mutation rates range over a wide range (ca.\ $0.5$--$23$).  Two loci appear to exhibit a mutation rates substantially higher than other loci, both of which appear to have high rates in the BP population as well (Figure \ref{fig:Kmar-BP-thetas-1}).

\begin{figure*}[h!]
\begin{centering}
\includegraphics[width=0.48\textwidth]{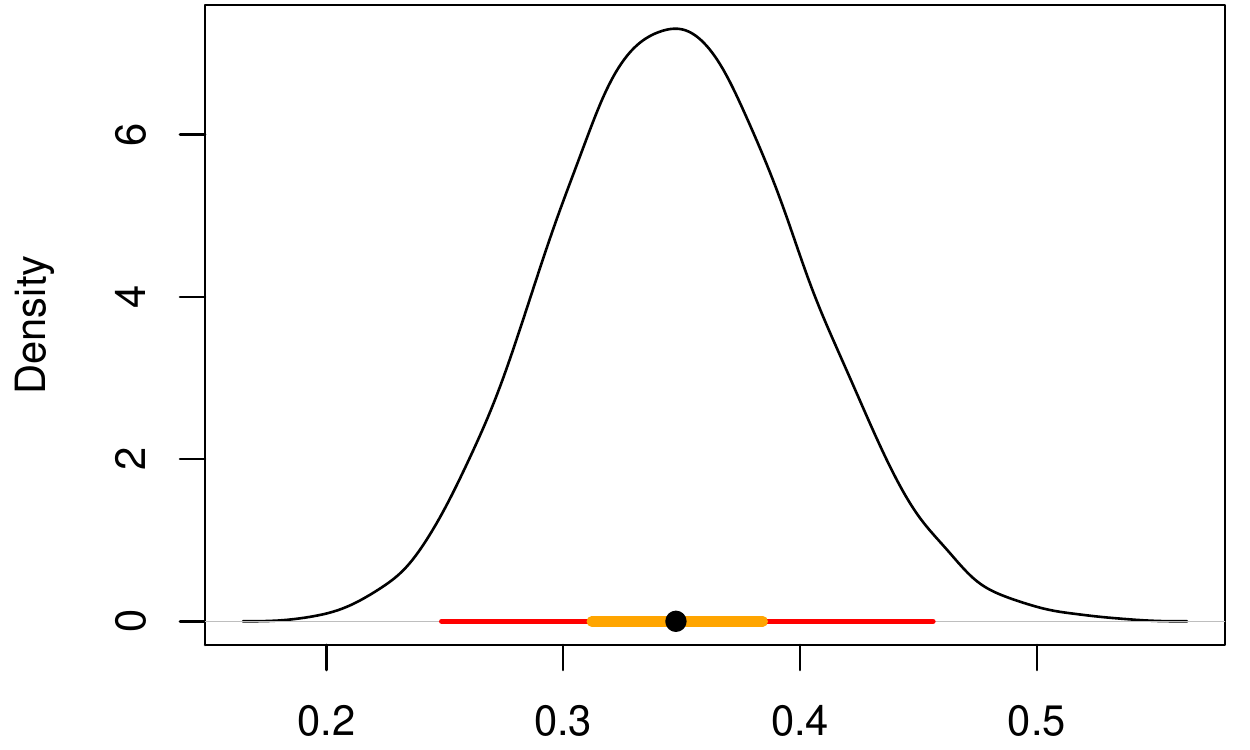}
\par\end{centering}
\protect\caption{\label{fig:Posterior-distributions-PG}Posterior distribution of
the uniparental proportion $s_A$ for the TC population. Also shown
are the 95\% BCI (red), 50\% BCI (orange), and median (black dot).}
\end{figure*}

All three methods confirm the inference of  \citet{Mackiewicz2006b} of much lower inbreeding in the TC population relative to the BP population. Our posterior distribution of uniparental proportion $s_A$
has a median and 95\% BCI of $0.35\,(0.25,0.45)$ (Figure \ref{fig:Posterior-distributions-PG}).
The median again lies between the $F_{IS}$-based estimate \eqref{fis} of $0.39$ and the \texttt{RMES} estimate of $0.33$, with its 95\%
CI of $(0.30,0.36)$.  In this case, \texttt{RMES} excluded from the analysis only a single locus, which was monomorphic in the sample.

\begin{figure}
\begin{centering}
\includegraphics[width=0.49\textwidth]{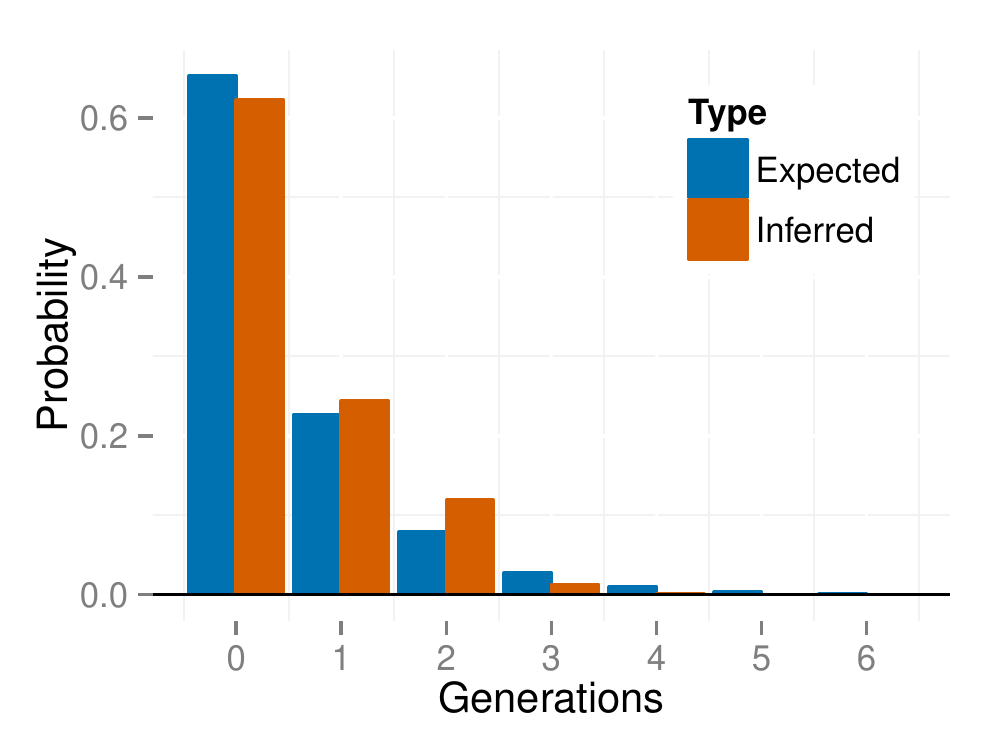}
\par\end{centering}
\protect\caption{\label{fig:Kmar-PG-t-dist-2-1-1}Empirical distribution of selfing
times $T$ across individuals, for \emph{K. marmoratus} (Population
TC). The histogram is averaged across posterior samples.}
\end{figure}

Figure \ref{fig:Kmar-PG-t-dist-2-1-1}
 shows the inferred distribution of the number of generations since the most recent outcross event ($T$) across
individuals, averaged over posterior uncertainty.  In contrast to
the BP population, the distribution of selfing time in the TC population appears to conform to the distribution expected under the inferred uniparental proportion ($s_A$), including a high fraction of biparental individuals ($T_k=0$).  Figure \ref{fig:Kmar-PG-t-dist-1-1} (Supplementary Material) presents the posterior distribution of the number of consecutive generations of selfing in the immediate ancestry of each individual.

\subsection*{Gynodioecious plant}

We next examine data from \emph{Schiedea salicaria}, a gynodioecious
member of the carnation family endemic to the Hawaiiian islands.  We analyzed genotypes at 9 microsatellite loci from 25 \emph{S. salicaria} individuals collected from west Maui and identified by \citet{WallaceCulleyWellerSakaiNepopkroeff2011} as non-hybrids.

\paragraph{Parameter estimation:}   Our gynodioecy model \eqref{paramgyno} comprises 4 basic parameters, including the relative seed set of females ($\sigma$) and the relative viability of uniparental offspring ($\tau$).  Our analysis of microsatellite data from the gynodioecious Hawaiian endemic \emph{Schiedea salicaria} \citep{WallaceCulleyWellerSakaiNepopkroeff2011} constrained the relative seed set of females to unity ($\sigma \equiv 1$), consistent with empirical results \citep{WellerSakai2005}.  In addition, we use results of experimental studies of inbreeding depression to develop an informative prior distribution for $\tau$:
\begin{equation}
\label{priortau}
\tau \sim \text{Beta}(2,8),
\end{equation}
the mean of which ($0.2$) is consistent with the results of greenhouse experiments reported by \citet{SakaiKarolyWeller1989}.

\citet{CampbellWellerSakai2010} reported a 12\% proportion of females ($n_f=27$ females among $n_{total}=221$ individuals).  As in the case of androdioecy \eqref{binmale}, we model this information by
\begin{equation}
\label{binfemale}
n_f \sim \text{Binomial}(n_\emph{total}, p_{f}),
\end{equation}
obtaining estimates from the extended likelihood function corresponding to the product of $\Pr(n_f | n_\emph{total}, p_f)$ and the likelihood of the genetic data.  We retain a uniform prior for the proportion of seeds of hermaphrodite set by self-pollen ($a$).

\paragraph{Results:}   Figure \ref{fig:Posterior-distributions-Schiedea} (Supplementary Material) presents posterior distributions of the basic parameters of the gynodioecy model \eqref{eq:gynodioecy}.  Our estimate of the uniparental proportion  $s_G$ (median $0.247$, 95\% BCI $(.0791,0.444)$)
is substantially lower than the $F_{IS}$-based estimate \eqref{fis} of $s_G=0.33$.  Although \texttt{RMES} excluded none of the loci, it gives an estimate of $s_G=0$, with a 95\% CI
of $(0,0.15)$.

Unlike the \emph{K. marmoratus} data sets, the \emph{S. salicaria} data set does not appear to provide substantial
evidence for large differences in locus-specific mutation rates across loci: Figure \ref{fig:Schiedea-thetas-1} (Supplementary Material) shows similar posterior
medians for across loci.

\begin{figure}[h!]
\begin{centering}
\includegraphics[width=0.5\textwidth]{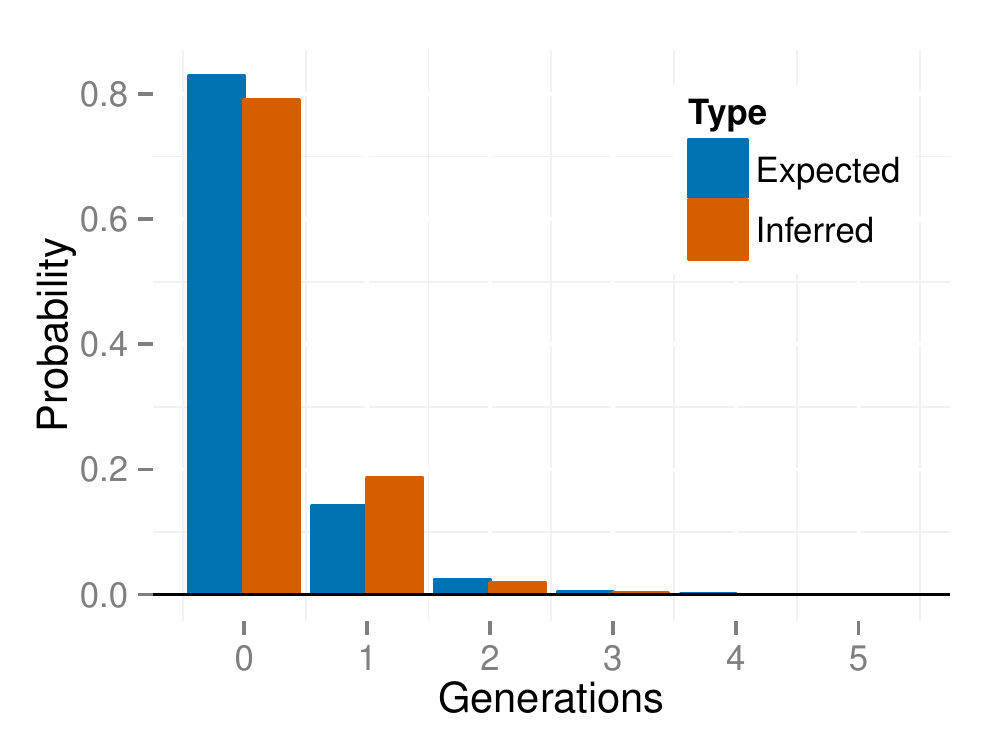}
\par\end{centering}

\protect\caption{\label{fig:Schiedea-t-dist-2-1}Empirical distribution of selfing
times $T$ across individuals, for \emph{S. salicaria}. The histogram
is averaged across posterior samples.}
\end{figure}

Figure \ref{fig:Schiedea-t-dist-2-1}
presents the inferred distribution of the number of generations since the most recent outcross event $T$ across
individuals, averaged over posterior uncertainty.   In contrast with the analysis of the \emph{K. marmoratus} BP population (Figure \ref{fig:Kmar-BP-t-dist-2-1}), the distribution appears to be consistent with the inferred uniparental proportion $s_G$.  Figure \ref{fig:Schiedea-t-dist-1} (Supplementary Material) presents the posterior distribution of the number of consecutive generations of selfing in the immediate ancestry of each individual.

Table \ref{tab:Parameter-estimates-1} presents posterior medians and 95\% BCIs for the proportion of uniparentals among reproductives ($s^{*}$), the proportion of seeds set by hermaphrodites by self-pollen ($a$), the  viability of uniparental offspring relative to biparental offspring ($\tau$), the proportion of females among reproductives ($p_{f}$), and the probability that a random gene derives from a female parent ($(1-s_G)F/2$). Comparison of the first (YYY) and
fifth (NYY) rows indicates that inclusion of the  genetic data more than doubles the
posterior median of $s^*$ (from $0.112$ to $0.247$) and
shrinks the credible interval.  Comparison of the first (YYY) and
third (YNY) rows indicates that counts of females and hermaphrodites greatly reduce the
posterior median of $p_f$ and accordingly change the proportional contribution of females to the gene pool ($(1-s_G)F/2$).  The bottom row of the table (NNN), showing a prior estimate for composite parameter 
$s^*$ of $0.0844\,(0.000797,0.643)$, illustrates that its induced prior
distribution departs from uniform on $(0,1)$, even though both of its components ($a$ and $\tau$) have uniform priors.
\newpage
\begin{landscape}
\begin{table*}[h!]
\protect\caption{\label{tab:Parameter-estimates-1}Parameter estimates for different
amounts of data. Estimates are given by a posterior median and a 95\%
BCI.}

\begin{tabular}{ccc|ccccc}
G & F & I & $s^{*}$ & $a$ & $\tau$ & $p_{f}$ & $(1-s_G)F/2$ \tabularnewline
\hline 
\hline 
Y & Y & Y & $0.247\,(0.0791,0.444)$ & $0.695\,(0.299,0.971)$ & $0.215\,(0.0597,0.529)$ & $0.125\,(0.0849,0.173)$ & $0.118\,(0.054,0.258)$\tabularnewline
\hline 
Y & Y & N & $0.267\,(0.0951,0.469)$ & $0.497\,(0.187,0.93)$ & $0.507\,(0.082,0.973)$ & $0.125\,(0.0851,0.174)$ & $0.0808\,(0.0398,0.191)$\tabularnewline 
\hline 
Y & N & Y & $0.213\,(0.045,0.402)$ & $0.742\,(0.379,1.00)$ & $0.252\,(0.0488,0.529)$ & $0.244\,(0.00,0.613)$ & $0.218\,(0.0,0.403)$\tabularnewline
\hline 
Y & N & N & $0.243\,(0.0608,0.429)$ & $0.628\,(0.268,0.999)$ & $0.611\,(0.167,1.00)$ & $0.354\,(0.00,0.072)$ & $0.223\,(0.00,0.394)$\tabularnewline
\hline 
N & Y & Y & $0.112\,(0.0026,0.588)$ & $0.496\,(0.0252,0.974)$ & $0.183\,(0.0277,0.513)$ & $0.125\,(0.0847,0.173)$ & $0.0956\,(0.0427,0.218)$\tabularnewline
\hline 
N & Y & N & $0.231\,(0.00391,0.776)$ & $0.504\,(0.025,0.973)$ & $0.493\,(0.0257,0.975)$ & $0.125\,(0.0847,0.173)$ & $0.0778\,(0.0392,0.172)$\tabularnewline
\hline 
N & N & Y & $0.0376\,(0.00,0.318)$ & $0.492\,(0.0122,0.957)$ & $0.0.185\,(0.00917,0.462)$ & $0.483\,(0.00,0.946)$ & $0.314\,(0.0361,0.500)$\tabularnewline
\hline 
N & N & N & $0.0844\,(0.000,0.643)$ & $0.497\,(0.0244,0.975)$ & $0.494\,(0.0252,0.975)$ & $0.479\,(0.0245,0.972)$ & $0.289\,(0.0313,0.5)$\tabularnewline
\hline 
\end{tabular}

Each row represents an analysis that includes (Y) or excludes (N) information, including genotype frequency data (G), counts of females (F), and replacement of the Uniform(0,1) prior on $\tau$ by an informative prior (I).
\end{table*}
\end{landscape}
\newpage

\section*{Discussion}

We introduce a model-based Bayesian method for the inference of the rate of self-fertilization and other aspects of a mixed mating system.  In anticipation of large (even genome-scale) numbers of loci, it uses the Ewens Sampling Formula (ESF) to determine likelihoods in a computationally efficient manner from frequency spectra of genotypes observed at multiple unlinked sites throughout the genome.  Our MCMC sampler explicitly incorporates the full set of parameters for each iconic mating system considered here (pure hermaphroditism, androdioecy, and gynodioecy), permitting insight into various components of the evolutionary process, including effective population size relative to the number of reproductives.

\subsection*{Assessment of the new approach}

\paragraph{Accuracy:} \citet{Enjalbert2000} and \citet{David2007} base estimates of selfing rate on the distribution of numbers of heterozygous loci.  Both methods strip genotype information from the data, distinguishing between only homozygotes and heterozygotes, 
irrespective of the alleles involved.  Loci lacking heterozygotes altogether (even if polymorphic) are removed from the analysis as uninformative about the magnitude of departure from Hardy-Weinberg proportions (Figure \ref{fig:Loci-and-data-unusable-by-RMES}).  As the observation of polymorphic loci with low heterozygosity provides strong evidence of inbreeding, exclusion of such loci by \texttt{RMES} \citep{David2007} may contribute to its loss of accuracy for high rates of selfing (Figure \ref{fig:error-posterior-median}).

Our method derives information from all loci.  Like most coalescence-based models, it accounts for the level of variation as well as the way in which variation is partitioned within the sample.  Even a locus monomorphic within a sample provides information about the age of the most recent common ancestor of the observed sequences, a property that was not widely appreciated prior to analyses of the absence of variation in a sample of human Y chromosomes \citep{DoritAkashiGilbert1995, FuLiYchr1996}.

Estimates of the rate of inbreeding produced by our method appear to show greater accuracy than \texttt{RMES} and the $F_{IS}$-based method \eqref{fis} over much of the parameter range (Figure \ref{fig:error-posterior-median}).  The increased error exhibited under very high rates of inbreeding ($s^\ast \approx 1$) may reflect violation of our assumption \eqref{eq:assumptions} that random outcrossing occurs on a much shorter time scale than mutation and coalescence.  Even though our method assumes that the rate of inbreeding lies in $(0,1)$, the posterior distribution for data generated under random outcrossing ($s^\ast =0$) does indicate greater confidence in low rates of inbreeding (Figure \ref{fig:Average-posterior-density-s0}).

Both \texttt{RMES} and our method invoke independence of genealogical histories of unlinked loci, conditional on the time since the most recent outcrossing event.  \texttt{RMES} seeks to approximate the likelihood by summing over the distribution of time since the most recent outcross event, but truncates the infinite sum at $20$ generations.  The increased error exhibited by \texttt{RMES} under high rates of inbreeding may reflect that the likelihood has a substantial mass beyond the truncation point in such cases.  Our method explicitly estimates the latent variable of time since the most recent outcross for each individual \eqref{eq:tlist}.  This quantity ranges over the non-negative integers, but values assigned to individuals are explored by the MCMC according to their effects on the likelihood.

\paragraph{Frequentist coverage properties:} Bayesian approaches afford a direct means of assessing confidence in parameter estimates, and our simulation studies suggest that the Bayesian Credible Intervals (BCIs) generated by our method have relatively
good frequentist coverage properties as well (Figure \ref{fig:coverage-BCI}).  The Confidence Intervals (CIs) reported by the maximum-likelihood  method \texttt{RMES} \citep{David2007} appear to perform less well (Figure \ref{fig:Comparison-RMES-CI}). Although \citet{David2007} describe \texttt{RMES} as determining CIs via the profile likelihood method \citep[see][]{Kreutz2013}, \texttt{RMES} holds constant parameters other than the uniparental proportion ($s^\ast$) instead of reoptimizing them to maximize the likelihood as $s^\ast$ varies.  The result is therefore not a true profile likelihood, which may explain the poor coverage properties of the CIs that \texttt{RMES} provides.

\paragraph{Model fit:}  Bayesian approaches also afford insight into the suitability of the underlying model.  Our method provides estimates of the number of generations since the most recent outcross event in the immediate ancestry of each individual ($T$).  We can pool such estimates of selfing times
to obtain an empirical distribution of the number of selfing generations, a procedure particularly useful for samples containing observation of the genotype of many individuals.  Under the assumption of a single population-wide rate of self-fertilization, we expect selfing time to have a geometric distribution with parameter corresponding to the estimated selfing rate.  Empirical distributions of the estimated number of generations since the last outcross appear consistent with this expectation for the data sets derived from the TC population of \emph{K. marmoratus}  (Figure \ref{fig:Kmar-PG-t-dist-2-1-1}) and from \emph{Schiedea} (Figure \ref{fig:Schiedea-t-dist-2-1}).  In contrast, the empirical distribution for the highly-inbred BP population of  \emph{K. marmoratus} (Figure \ref{fig:Kmar-BP-t-dist-2-1}) shows an absence of individuals formed by random outcrossing ($T=0$).  That our method accurately estimates $T$ from simulated data (Figure \ref{fig:reconstructed-selfing-times-1}) argues against attributing the inferred deficiency of biparental individuals in the BP data set to an artifact of the method.  Rather, the deficiency may indicate a departure from the underlying model, which assumes reproduction only through self-fertilization or random outcrossing.  In particular, biparental inbreeding as well as selfing may reduce the fraction of individuals formed by random outcrossing.  Mis-scoring of heterozygotes as homozygotes due to null alleles or other factors, a possibility directly addressed by \texttt{RMES} \citep{David2007}, may also in principle  contribute to the paucity of outbred individuals.

\subsection*{Components of inference}

\paragraph{Locus-specific mutation rates:}   Our method estimates the scaled mutation rate \eqref{rates} at each locus using the Dirichlet Process Prior (DPP).  This approach improves on existing methods in several ways.  First, we estimate a single parameter for each locus instead of estimating multiple allele frequencies per locus as do \citet{Enjalbert2000}.  
Second, we estimate for each locus the scaled mutation rate, a fundamental component of the evolutionary process, rather than the heterozygosity \eqref{apphet}, a random outcome of that process.  Third, incorporation of the DPP permits the simultaneous estimation of the number of classes of mutation rates, the mutation rate for each class, and the class membership of each locus.  It accords the increased accuracy derived from pooling loci with similar mutation rates without  \emph{a priori} knowledge of the partitioning of loci among rate classes or even the number of classes.

\paragraph{Joint inference of mutation and inbreeding rates:}  For the infinite-alleles model of mutation, the Ewens Sampling Formula \citep[ESF,][]{Ewens1972} provides the probability of any allele frequency spectrum (AFS) observed at a locus in a sample derived from a panmictic population.  Under partial self-fertilization, the ESF provides the probability of an AFS observed among genes, each sampled from a distinct individual.  For such genic (as opposed to genotypic) samples, the coalescence process under inbreeding is identical to the standard coalescence process, but with a rescaling of time \citep{Fu1997, Nordborg1997}.  Accordingly, genic samples may serve as the basis for the estimation of the single parameter of the ESF, the scaled mutation rate $\theta^\ast$ \eqref{eq:theta-star}, but not the rate of inbreeding apart from the scaled mutation rate.

Our method uses the information in a genotypic sample, the genotype frequency spectrum, to infer both the uniparental proportion $s^\ast$ and the scaled mutation rate $\theta^\ast$.   Our sampler reconstructs the genealogical history of a sample of diploid genotypes only to the point of the most recent random-outcross event of each individual, with the number of consecutive generations of inbreeding in the immediate ancestry of a given individual ($T_k$ for individual $k$) corresponding to a latent variable in our Bayesian inference framework.  Invocation of the ESF beyond the point at which all lineages reside in separate individuals obviates the necessity of further genealogical reconstruction.  As a consequence, our method may be better able to accommodate  genome-scale magnitudes of observed loci ($L$).

Identity disequilibrium \citep{Cockerham1968}, the correlation in heterozygosity across loci within individuals, reflects that all loci within an individual experience the most recent random-outcross event at the same time, irrespective of physical linkage.  The heterozygosity profile of individual $k$ provides information about $T_k$ \eqref{eq:T_k}, which in turn reflects the uniparental proportion $s^\ast$.  Observation of multiple individuals provides a basis for inference of both the uniparental proportion $s^\ast$ and the scaled mutation rate $\theta^\ast$.

\paragraph{Identifiability:} In an analysis based solely on the genotype frequency spectrum observed in a sample, the likelihood depends on just two composite parameters:  the probability that a random individual is uniparental ($s^\ast$) and the scaled rates of mutation $\boldsymbol{\Theta}^\ast$ \eqref{eq:bigTheta} across loci.  Any model for which the parameter set $\mathbf{\Psi}$ \eqref{likeparameters} comprises more than one parameter is not fully identifiable from the genetic data alone.   In the pure hermaphroditism model \eqref{eq:pure-h}, for example, basic parameters $\tilde s$ (fraction of fertilizations by selfing) and $\tau$ (relative viability of uniparental offspring) are nonidentifiable:  any assignments that determine the same values of composite parameters $s^\ast$ and $\boldsymbol{\Theta}^\ast$ have the same likelihood.

For each basic parameter in $\mathbf{\Psi}$ beyond one, identifiability requires incorporation of additional information beyond the genetic data.  A full treatment of such information requires expansion of the likelihood function to encompasses an explicit model of the new information.  Our androdioecy model \eqref{eq:androdioecy}, for example, comprises 3 parameters, including the frequency of males among reproductives ($p_m$) as well as $\tilde s$ and $\tau$.  In our analysis of microsatellite data from the killifish \emph{Kryptolebias marmoratus} \citep{Mackiewicz2006b, TatarenkovEarleyTaylorAvise2012}, the expanded likelihood function corresponds to the product of the probability of the genetic data and the probability of the number of males observed among a total number of individuals \eqref{binmale}.  In the absence of information regarding inbreeding depression ($\tau$), we assigned $\tau \equiv 1$ to permit estimation of the uniparental proportion ($s^\ast$) under a uniform prior distribution.  This assignment does not affect the reliability of our estimates ($s^\ast$, $\boldsymbol{\Theta}^\ast$, $p_m$, $S_A$, \emph{etc}.); rather, the analysis is agnostic concerning the influence of the relative viability of inbred offspring ($\tau$) and the rate of self-fertilization ($\tilde s$) in determining the probability that a random individual is uniparental ($s^\ast$).

Non-identifiable parameters can also be estimated through the incorporation of informative priors.  Because identifiability is defined in terms of the likelihood, which is unaffected by priors, such parameters remain non-identifiable.  Even so, informative priors assist in their estimation through Bayesian approaches, which do not require parameters to be identifiable.  To explore the data set from \emph{Schiedea salicaria} \citep{WallaceCulleyWellerSakaiNepopkroeff2011}, we use our 4-parameter gynodioecy model \eqref{eq:gynodioecy}, the basic parameters of which include the proportion of females among reproductives ($p_f$), the relative seed set of females ($\sigma$), the relative viability of uniparental offspring ($\tau$), and the proportion of seeds of hermaphrodites set by self-pollen ($a$).  In a manner similar to the androdioecy study, our analysis uses an extended likelihood function, modeling the number of females as a binomial random variable \eqref{binfemale}.  In addition, we use earlier experimental evidence to justify the assignment of $\sigma \equiv 1$ \citep{WellerSakai2005} and to develop an informative prior for $\tau$ \citep[\eqref{priortau}:][]{SakaiKarolyWeller1989}.  This procedure permits estimation of 3 basic parameters,
including the proportion of seeds of hermaphrodites set by self-pollen ($a$).

\subsection*{Beyond estimation of the selfing rate}

Our MCMC implementation updates the full set of basic parameters, with likelihoods determined from the implied values of composite parameters $s^\ast$ and $\boldsymbol{\Theta}^\ast$.  Incorporation of additional information, either through extension of the likelihood or through informative priors, permits inference not only of the basic parameters but also of functions of the basic parameters.   For example, Table \ref{tab:Parameter-estimates-1} includes estimates of the proportion of seeds of hermaphrodites set by self-pollen ($a$) and the probability that a random gene derives from a female parent ($(1-s_G)F/2$) in gynodioecious \emph{S. salicaria}.  We are not aware of other studies in which these quantities have been inferred from the pattern of neutral genetic variation observed in a random sample.

\begin{figure}[h!]
\begin{centering}
\includegraphics[width=\textwidth]{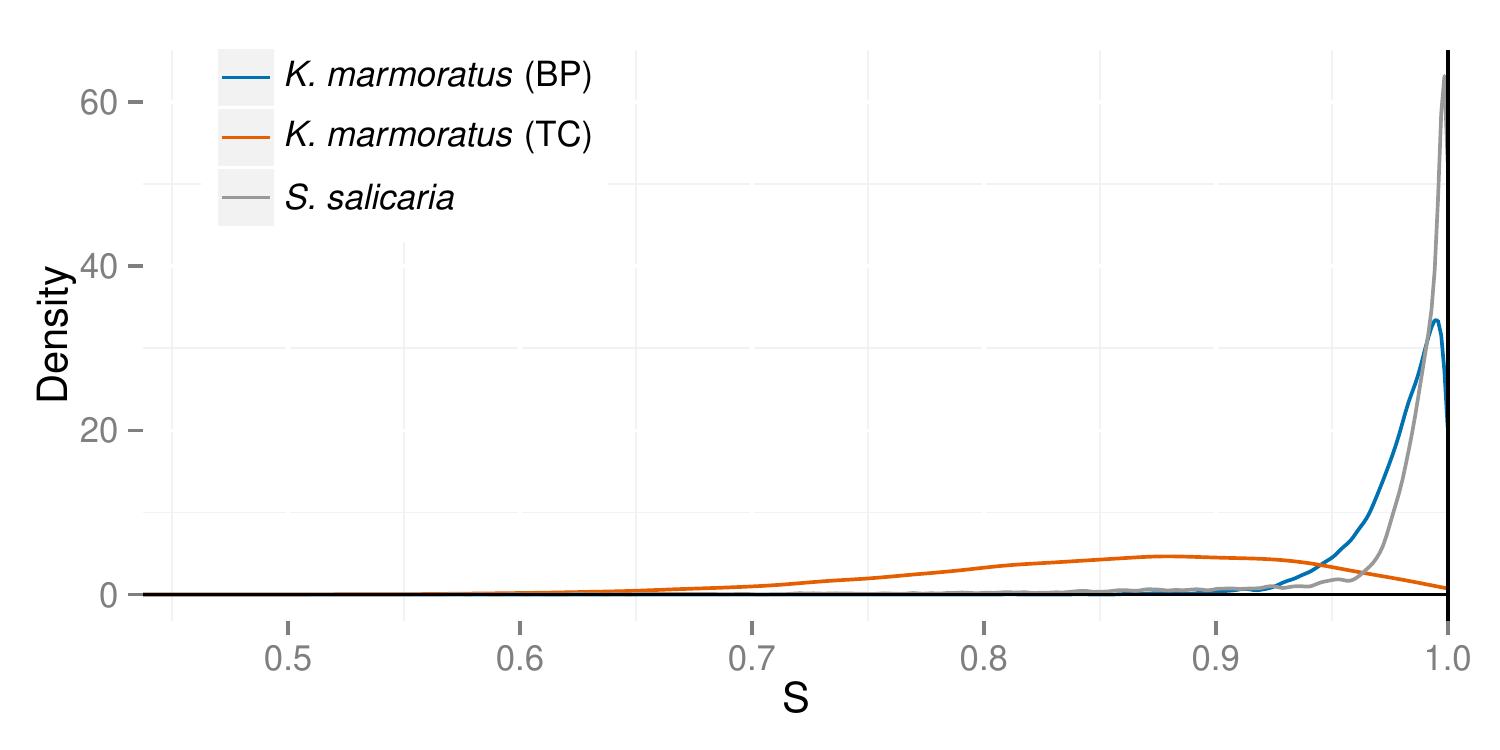}
\par\end{centering}

\protect\caption{\label{combinedS}Posterior distributions of relative effective number $S$ \eqref{rates} for data sets derived from \emph{Kryptolebias marmoratus} (BP and TC populations) and \emph{Schiedea salicaria}.}
\end{figure}

Among the most biologically-significant functions to which this approach affords access is relative effective number $S$ \eqref{rates}, a fundamental component of the reproductive value of the sexes \citep{Fisher1958}.  We denote the probability that a pair of genes, randomly drawn from distinct individuals, derive from the same parent in the preceding generation as the rate of parent-sharing ($1/N^\ast$).  Its inverse ($N^\ast$) corresponds to the ``inbreeding effective size'' of \citet{CrowDenniston88}.  Relative effective number $S$ is the ratio of $N^\ast$ to the total number of reproductive individuals.  For example, in the absence of inbreeding ($s^\ast=0$), $N^\ast$ in our gynodioecy model \eqref{eq:gynodioecy} corresponds to Wright's (\citeyear{Wright69}) harmonic mean expression for effective population size and $S$ to the ratio of $N^\ast$ and $N_f+N_h$, the total number of reproductive females and hermaphrodites.  In general ($s^\ast \ge 0$), relative effective size $S$ reflects reductions in effective size due to inbreeding in addition to differences in numbers of the sexual forms.  Figure \ref{combinedS} presents posterior distributions of $S$ for the 3 data sets explored here.  These results suggest that relative effective number $S$ in each of the natural populations surveyed lies close to its maximum of unity, with the effective number defined through the rate of parent-sharing approaching the total number of reproductives.

\section*{Acknowledgments}
This project was initiated during a sabbatical visit of MKU to the Department of Ecology and Evolutionary Biology at the University of California at Irvine.  We thank Francisco J. Ayala for exceedingly gracious hospitality, and Diane R. Campbell and all members of the Department for stimulating interactions.  We thank Lisa E. Wallace for making available to us microsatellite data from \emph{Schiedea salicaria}.  Public Health Service grant GM 37841 (MKU) provided partial funding for this research.

\bibliographystyle{genetics}
\bibliography{coalescent}

\titleformat{\section}{\Large\bf}{Appendix \thesection}{0.5em}{#1}[]

\begin{appendices}
\numberwithin{equation}{section}

\section{The last-sampled gene}\label{nextsampledgene}

We address the probability that the last-sampled gene in a sample of size $i$ represents a novel allele \eqref{novelpi}.

Under the infinite alleles model of mutation, a single mutation in a lineage suffices to distinguish a new allele.  We denote the last-sampled gene in a sample of size $i$ as the focal gene, and consider the level of the genealogical tree in which its ancestral lineage either receives a mutation or joins the gene tree of the sample at size $(i-1)$.  Level $l$ of the entire ($i$-gene) gene tree corresponds to the segment in which $l$ lineages persist.

The probability that the line of descent of the focal gene terminates in a mutation immediately, in level $i$ of the genealogy, is
\begin{equation*}
\frac{u}{nu + \binom{i}{2}/N^\ast} =\frac{\theta^\ast}{i(\theta^\ast+i-1)}.
\end{equation*}
In general, the probability that the lineage of the focal gene terminates on level $l>2$ is
\begin{gather*}
\frac{(i-1)u + \binom{i-1}{2}/N^\ast}{iu+\binom{i}{2}/N^\ast} \; \frac{(i-2)u+\binom{i-2}{2}/N^\ast}{(i-1)u + \binom{i-1}{2}/N^\ast}\; \ldots \; \frac{lu+\binom{l}{2}/N^\ast}{(l+1)u + \binom{l+1}{2}/N^\ast}\; \frac{u}{lu + \binom{l}{2}/N^\ast}\\
=\frac{\theta^\ast}{i(\theta^\ast+i-1)}.
\end{gather*}
This expression illustrates the invariance over termination orders noted by \citet{GriffithsLessard05}.  Summing over all levels, including level 2, for which a mutation in either remaining lineage ensures that the focal gene represents a novel allele, we obtain the overall probability that the last-sampled gene represents a novel allele:
\begin{equation*}
\frac{\theta^\ast(i-2)}{i(\theta^\ast+i-1)}+\frac{2\theta^\ast}{i(\theta^\ast+i-1)}=\frac{\theta^\ast}{\theta^\ast+i-1}.
\end{equation*}

\section{Estimators of $\boldsymbol{F_{IS}}$}\label{appfis}

We follow \citet{Weir1996} in developing an estimate of the uniparental proportion $s^\ast$ from $F_{IS}$ alone \eqref{fis}.

For a single locus, a simple estimator of $F_{IS}$  corresponds to
\begin{equation*}
\widehat{F_{IS}} = 1-\frac{O}{E},
\end{equation*}
for $O$ the observed fraction of heterozygotes in the sample
and $E$ the expected fraction based on Hardy-Weinberg proportions given the observed allele frequencies.  Explicitly, we have
\begin{equation*}
\widehat{F_{IS}} =1-\frac{1-\sum_{u}\tilde{P}_{uu}}{1-\sum_{u}\tilde{p}_{u}^{2}}= \frac{\left(\sum_{u}\tilde{P}_{uu}-\tilde{p}_{u}^{2}\right)}{1-\sum_{u}\tilde{p}_{u}^{2}},
\end{equation*}
for $\tilde{p}_u$ the frequency of allele $u$ in the sample and 
$\tilde{P}_{uu}$ the frequency of homozygous genotype $uu$ in the sample.
However, this estimator can be substantially biased for small samples, leading to underestimation of $F_{IS}$  \citep{Weir1996}.

To address this bias and accommodate multiple loci, we instead adopt
\begin{equation}
\label{good_fis}
\widehat{F_{IS}} = \frac{\sum_{l=1}^{L}\left[\sum_{u=1}^{K_{l}}\left(\tilde{P}_{luu}-\tilde{p}_{lu}^{2}\right)+\left(1-\sum_{u=1}^{K_{l}}\tilde{P}_{luu}\right)/2n \right]}{\sum_{l=1}^{L}\left[\left(1-\sum_{u=1}^{K_{l}}\tilde{p}_{lu}^{2}\right)-\left(1-\sum_{u=1}^{K_{l}}\tilde{P}_{luu}\right)/2n\right]},
\end{equation}
for $n$ the number of diploid genotypes observed, $L$ the number of loci, and $K_l$ the number of alleles at locus $l$.
While this estimator is also biased in general, it corresponds to the ratio of unbiased estimators of $F_{IS}\cdot\sum_{l}\left(1-\sum_{u}p_{lu}^{2}\right)$
and $\sum_{l}\left(1-\sum_{u}p_{lu}^{2}\right)$, in which
$p_{lu}$ is the frequency of allele $u$ at locus $l$ in the entire
population \citep{Weir1996}.  Our analysis of simulated data (Appendix \ref{sec:simdata}) indicates that this estimator is more accurate than an estimator that simply averages single-locus
estimates:
\begin{equation}
\label{averaged_fis}
\widehat{F_{IS}} = \frac{1}{L}\sum_{l=1}^{L}\frac{\sum_{u=1}^{K_{l}}\left(\tilde{P}_{luu}-\tilde{p}_{lu}^{2}\right)+\left(1-\sum_{u=1}^{K_{l}}\tilde{P}_{luu}\right)/2n}{\left(1-\sum_{u=1}^{K_{l}}\tilde{p}_{lu}^{2}\right)-\left(1-\sum_{u=1}^{K_{l}}\tilde{P}_{luu}\right)/2n}.
\end{equation}
Our $F_{IS}$-based estimates \eqref{fis} incorporate \eqref{good_fis} and not \eqref{averaged_fis}.

\section{Implementation of the MCMC}

\paragraph{State space:} The state space for the Markov chain of our MCMC sampler includes times across sampled individuals since the last outcross event $\mathbf{T}$ \eqref{eq:tlist}, coalescence events across individuals and loci since that event $\mathbf{I}$ \eqref{eq:ilist}, and model-specific parameters $\mathbf{\Psi}$ \eqref{likeparameters}.  The state space also comprises the scaled mutation rates $\mathbf{\Theta}^{*}$ \eqref{eq:bigTheta}, which are determined by $\mathbf{C}$, a list specifying the mutation rate category $C_l$ for locus $l=1\ldots L$, and $\mathbf{Z}$, a list specifying the scaled mutation rate $Z_{i}$ for category $i=1\ldots L+4$.  In particular, the scaled mutation rate at locus $l$ corresponds to
\begin{equation}
\label{mutz}
\theta_{l}^{*} = Z_{C_{l}}.
\end{equation}
At any given point in the MCMC, the state of the Markov chain corresponds to $(\mathbf{I},\mathbf{T},\mathbf{\Psi},\mathbf{C},\mathbf{Z})$.

\paragraph{Iterations:}
Each iteration of our MCMC sampler performs multiple updates, with
each variable updated at least once per iteration. We recorded the state sampled by the MCMC at each iteration.  For analyses
of simulated data sets, we ran Markov chains for 2000 iterations,
discarding the first 200 iterations as burn-in. For analyses of the actual
data sets, we ran Markov chains for 100,000 iterations, discarding the
first 10,000 iterations as burn-in.  Convergence appeared to occur as rapidly for actual data as for simulated data, but we found empirically that the larger number of samples were needed to achieve smooth density plots for the actual data sets.

\paragraph{Transition kernels:} Updating of the continuous variables of mutation
rates $\left\{ Z_{l}\right\}$ \eqref{mutz} and model-specific parameters $\mathbf{\Psi}$ \eqref{likeparameters} uses both Metropolis-Hastings (MH)
transition kernels and auto-tuned slice-sampling transition kernels. Updating of the discrete variables $\left\{ C_{l}\right\}$ uses a Gibbs
transition kernel.

\paragraph{Efficient inference on selfing times through collapsed Metropolis-Hastings:} Simple Metropolis-Hastings (MH) proposals that separately update the time since the most recent outcross event ($T_{k}$) and coalescence history since that event ($I_{\cdot k}$) lead to extremely
poor mixing efficiency.  Strong correlations between $T_{k}$
and $I_{\cdot k}$ cause changes to $T_{k}$ to be rejected with high probability unless $I_{\cdot k}$
is updated as well. For example, consider proposing a change of $T_{k}$
from $1$ to $0$. When $T_{k}=1$, on average $I_{lk}$
will be $1$ at half of the loci and $0$ at the remaining loci. 
If any of the $I_{lk}=1$, a move to $T_{k}=0$ will always be rejected because the probability of a coalescence event more recently than the most recent outcross event is $0$ if the sampled individual is itself a product of outcrossing.  To permit acceptance of changes to $T_{k}$, we introduce
a proposal for $T_{k}$ that also changes $I_{\cdot k}$.

The scheme starts from the value $T_{k}=t_{k}$ and proposes a new
value $t_{k}^{\prime}$. In standard MH within Gibbs, we would compute
the probability of $T_{k}=t_{k}$ and of $T_{k}=t_{k}^{\prime}$ given
that all other parameters are unchanged. We modify this MH scheme
to compute probabilities without conditioning on the coalescence indicators
for individual $k$. However, the coalescence indicators for other
individuals are still held constant. To compute this probability,
let $J$ indicate all the coalescence indicators $I_{\cdot y}$ where
$y\ne k$. Then 
\begin{eqnarray*}
\Pr(\mathbf{X},\mathbf{T},\mathbf{J},s,\theta) & = & \Pr(\mathbf{X},\mathbf{J}|\mathbf{T},s,\theta)\Pr(\mathbf{T}|s)\Pr(s)\Pr(\theta).
\end{eqnarray*}
We introduce $\mathbf{I}_{\cdot k}$ by summing over all possible
values $\mathbf{i}_{\cdot k}$.
\begin{eqnarray*}
\Pr(\mathbf{X},\mathbf{J}|\mathbf{T},s,\theta) & = & \sum_{i_{\cdot k}}\Pr(\mathbf{X},\mathbf{I}_{\cdot k}=\mathbf{i}_{\cdot k},\mathbf{J}|\mathbf{T},s,\theta).
\end{eqnarray*}
 Since the $i_{lk}$ for different loci are independent given $T_{k}$,
we have

\begin{eqnarray*}
\Pr(\mathbf{X},\mathbf{J}|\mathbf{T},s,\theta) & = & \sum_{i_{\cdot k}}\prod_{l=1}^{L}\Pr(\mathbf{X}_{l},I_{lk}=i_{lk},\mathbf{J}_{l}|\mathbf{T},s,\theta)\\
 & = & \prod_{l=1}^{L}\sum_{i_{lk}}\Pr(\mathbf{X}_{l},\mathbf{I}_{lk}=i_{lk},\mathbf{J}_{l}|\mathbf{T},s,\theta).
\end{eqnarray*}
Therefore, for specific values of $\mathbf{T}$ and $\mathbf{J}$,
we can compute the sum over all possible values of $\mathbf{I}_{\cdot k}$
for $l=1\ldots L$ in computation time proportional to $L$ instead
of $2^{L}$. This is possible because the $L$ coalescence indicators
for individual $k$ each affect different loci, and are conditionally
independent given $T_{k}$ and $\mathbf{J}$. 

After accepting or rejecting the new value of $T_{k}$ with $I_{\cdot k}$
integrated out, we must choose new values for $\mathbf{I}_{\cdot k}$
given the chosen value of $T_{k}$. Because of their conditional independence,
we may separately sample each coalescence indicator $I_{lk}$ for
$l=1\ldots L$ from its full conditional given the chosen value of
$T_{k}$. This completes the collapsed MH proposal.

\section{Analysis of simulated data}\label{sec:simdata}

\paragraph{Simulations:} Our simulator (\url{https://github.com/skumagai/selfingsim}) was developed using \texttt{simuPOP}, publicly available at \url{http://simupop.sourceforge.net/}.  It explicitly represents $N=10,000$ individuals, each bearing two genes at each of $L$ unlinked loci.  Mutations arise at locus $l$ at scaled rate $\theta_l$ \eqref{rates}, in accordance with the the infinite-alleles model.

We assigned to uniparental proportion $s^{*}$ values ranging from $0.01$ to $0.99$, with half of the $L=32$ loci assigned scaled mutation rate $\theta=0.5$ and the remaining loci $\theta=1.5$.

We conducted $10^2$ independent simulations for each assignment of $s^{*}$.  Each simulation was initialized with each of the $2N \times 32$ genes representing a unique allele.  Most of this maximal heterozygosity was lost very rapidly, with allele number and allele frequency spectrum typically stabilizing well within $10N$ generations.  After $20N$ generations, we recorded the realized population, from which 100 independent samples of $L=32$ loci of size $n=70$ were extracted.  From this collection, we randomly chose $L=6$ loci and subsampled 100 independent samples of size $n=6$.

\paragraph{Analysis:} To $10^2$ independent samples from each of $10^2$ independent simulations for each assignment of the uniparental proportion $s^\ast$, we applied our Bayesian method, the $F_{IS}$ method, and \texttt{RMES}.  
Our Bayesian method is open-source and can be obtained at
\begin{flushleft}
\url{https://github.com/bredelings/BayesianEstimatorSelfing/}.
\end{flushleft}
We used the implementation of \texttt{RMES} \citep{David2007} provided at
\begin{flushleft}
\url{http://www.cefe.cnrs.fr/images/stories/DPTEEvolution/Genetique/fichiers\%20Equipe/RMES\%202009\%282\%29.zip}.
\end{flushleft}

\end{appendices}

\newpage

\renewcommand{\textfraction}{0.01}
\renewcommand{\topfraction}{0.01}
\renewcommand{\bottomfraction}{0.01}
\renewcommand{\floatpagefraction}{0.01}

\setcounter{totalnumber}{1}
\setcounter{equation}{0}
\setcounter{section}{0}
\setcounter{figure}{0}
\setcounter{page}{2}
\renewcommand{\thefigure}{S\arabic{figure}}

\fancypagestyle{supplement}{%
    \renewcommand{\headrulewidth}{0pt}%
    \fancyhf{}%
    \fancyfoot[C]{B. D. Redelings et al.}%
    \fancyfoot[R]{\bf \thepage\ SI}%
}

\pagestyle{supplement}
\noindent
\begin{centering}
\textbf{File S1}\\
\textbf{Supplementary Methods}\\
\end{centering}

\titleformat{\section}{\Large\bf}{\thesection}{0.5em}{#1}[]

\section{Indicators of accuracy}

To compare the accuracy of our Bayesian method to \texttt{RMES} and the $F_{IS}$ method, which produce point estimates,  we summarize the posterior distribution of the uniparental proportion $s^\ast$ by the median.  Here, we compare the median to the mode and mean of the posterior distribution.

\begin{figure}[H!]
\begin{centering}
\subfloat[$n=10$, $L=6$]{\begin{centering}
\includegraphics[width=0.48\textwidth]{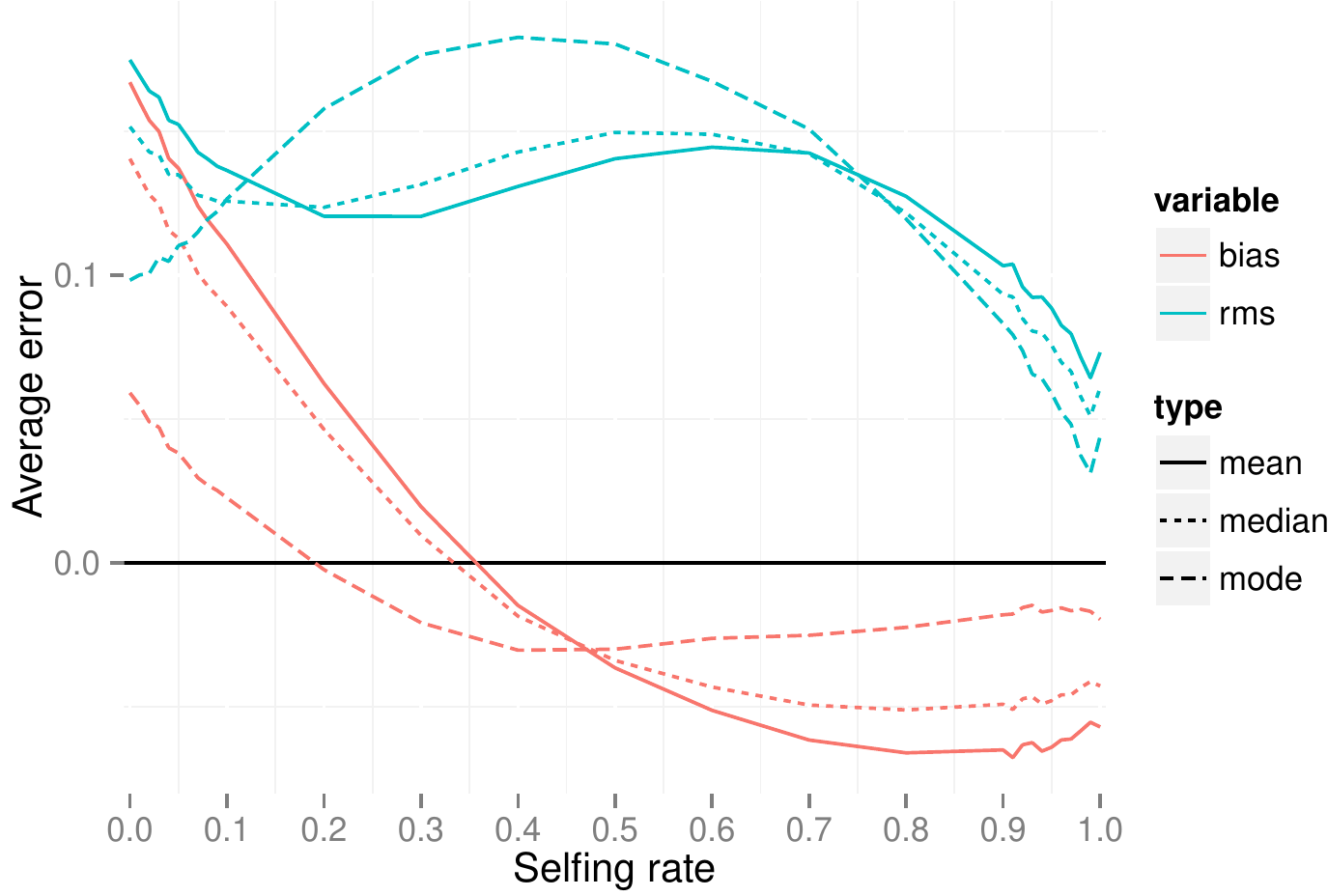}
\par\end{centering}

}\subfloat[$n=70$, $L=32$]{\begin{centering}
\includegraphics[width=0.48\textwidth]{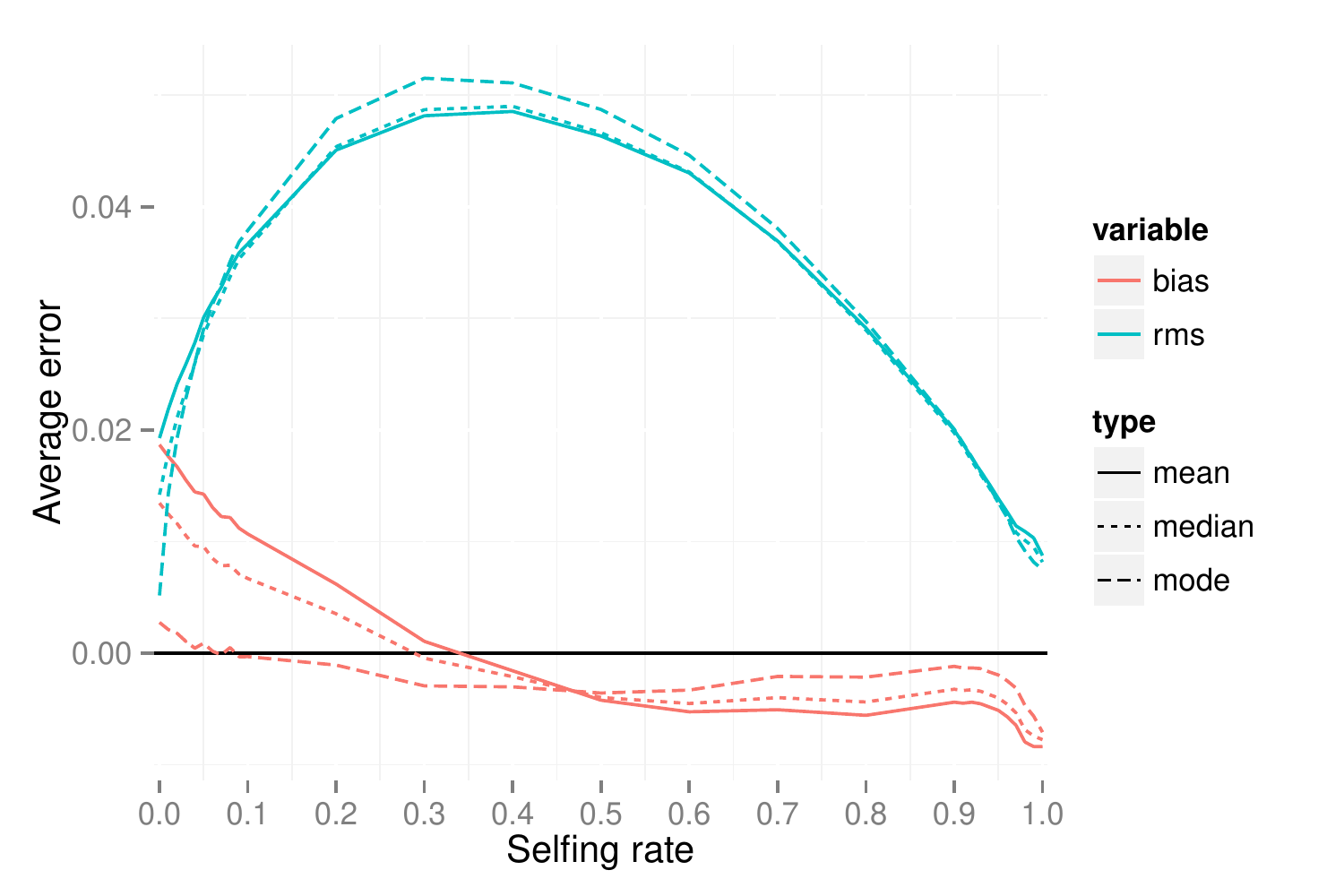}
\par\end{centering}

}
\par\end{centering}

\protect\caption{\label{fig:error-posteriors}Errors for the posterior mean, posterior
median, and posterior mode. Blue curves (rms) indicate the root-mean-squared error,
and red curves (bias) the average deviation. Averages are taken across
simulated data sets at each true value of the selfing rate $s^{*}$.}
\end{figure}

Figure \ref{fig:error-posteriors} suggests that the bias and root-mean-squared (rms) error of these three indices
exhibit different properties.  For example, the posterior mode shows smaller bias throughout the parameter range, but the median and mean show smaller rms error for $s^{*}$ near the boundaries (near $0$ or $1$).

\section{Average error}

\begin{figure}[H!]
\begin{centering}
\includegraphics[width=\textwidth]{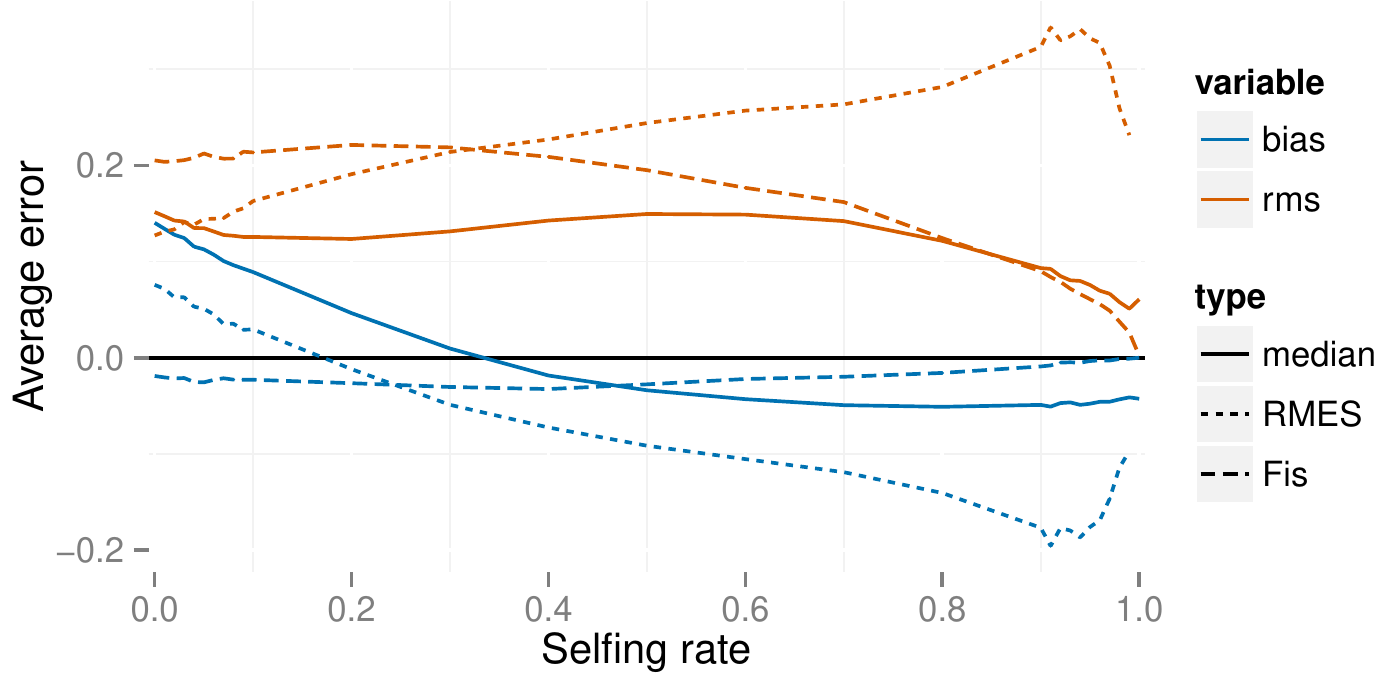}
\par\end{centering}
\protect\caption{\label{fig:error-posterior-median-small}Errors for the full likelihood
(posterior median), \texttt{RMES}, and $F_{IS}$ methods for a small sample ($n=10$ individuals, $L=6$ loci). In the legend, rms indicates the
root-mean-squared error and bias the average deviation.
Averages are taken across simulated data sets at each true value of
$s^{*}$.}
\end{figure}

As for the case of large simulated data sets (Figure \ref{fig:error-posterior-median}), Figure \ref{fig:error-posterior-median-small} indicates that upon application to smaller samples ($n=10$ individuals, $L=6$ loci), both \texttt{RMES} and our method show positive bias upon application to data sets for which the true uniparental proportion $s^{*}$ is close to zero and negative bias for $s^{*}$ 
close to unity.  It further indicates that while both methods exhibit more error for small samples than large samples, our Bayesian method exhibits less error than \texttt{RMES} throughout the range of the uniparental proportion ($s^\ast$).

\section{Frequentist coverage}

\begin{figure*}[H!]
\begin{centering}
\includegraphics[width=0.48\textwidth]{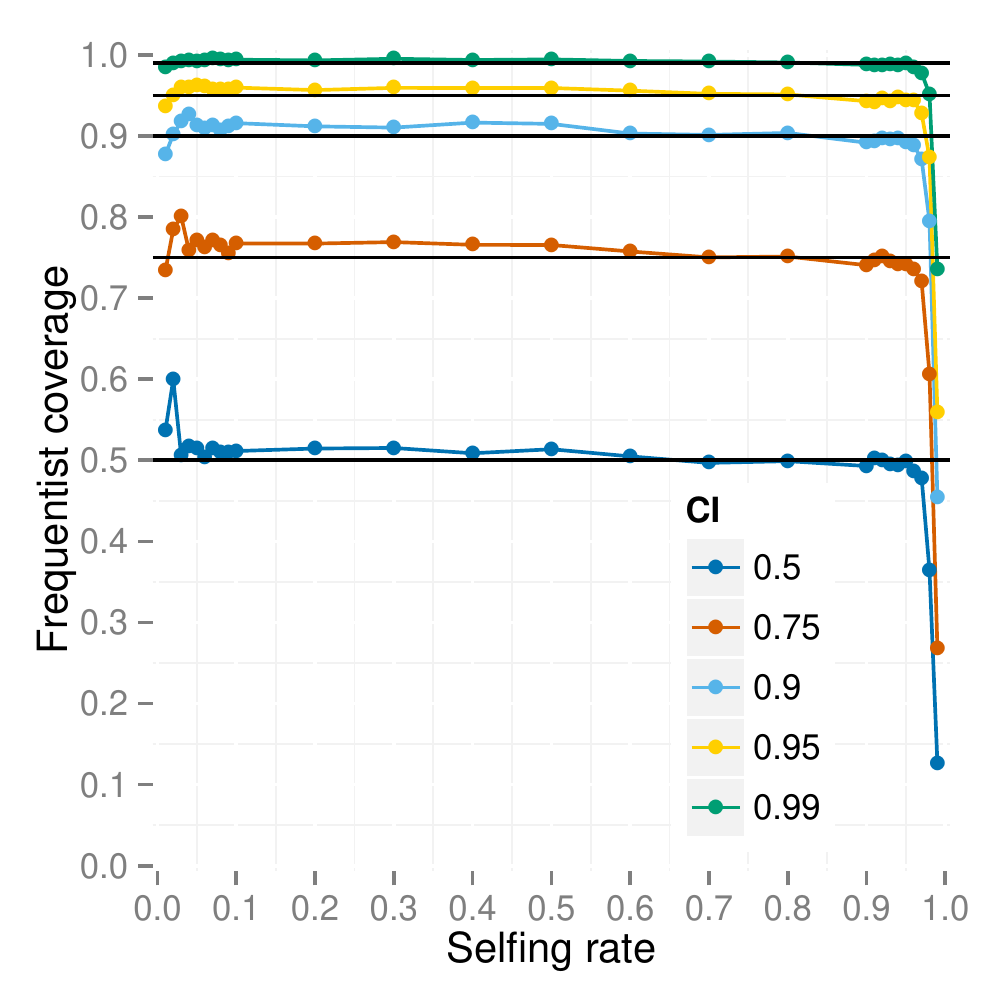}
\par\end{centering}
\protect\caption{\label{fig:coverage-BCI}Frequentist coverage for Bayesian credible
intervals at different levels of credibility under the large sampling regime ($n=70, L=32$).}
\end{figure*}

As for the 95\% BCIs (Figure \ref{fig:Comparison-RMES-CI}),
Figure \ref{fig:coverage-BCI} 
indicates that BCIs of different nominal values ($0.5$, $0.75$, $0.9$, $0.95$, and $0.99$)
display the same pattern, with coverage exceeding the
desired value for intermediate true $s^{*}$ values and dipping
below the desired value for very high values of $s^{*}$.
Coverage is closer to the nominal value for the $0.99$ and $0.95$
levels than for the $0.5$ level.

\section{Data analysis}

\subsection{Androdioecious vertebrate}

\paragraph{Low outcrossing rate:} As noted in the main text, we find evidence of a multimodal distribution of mutation rates in the BP population of \emph{K. marmoratus}.
\begin{figure*}[H!]
\begin{centering}
\includegraphics[width=1\textwidth]{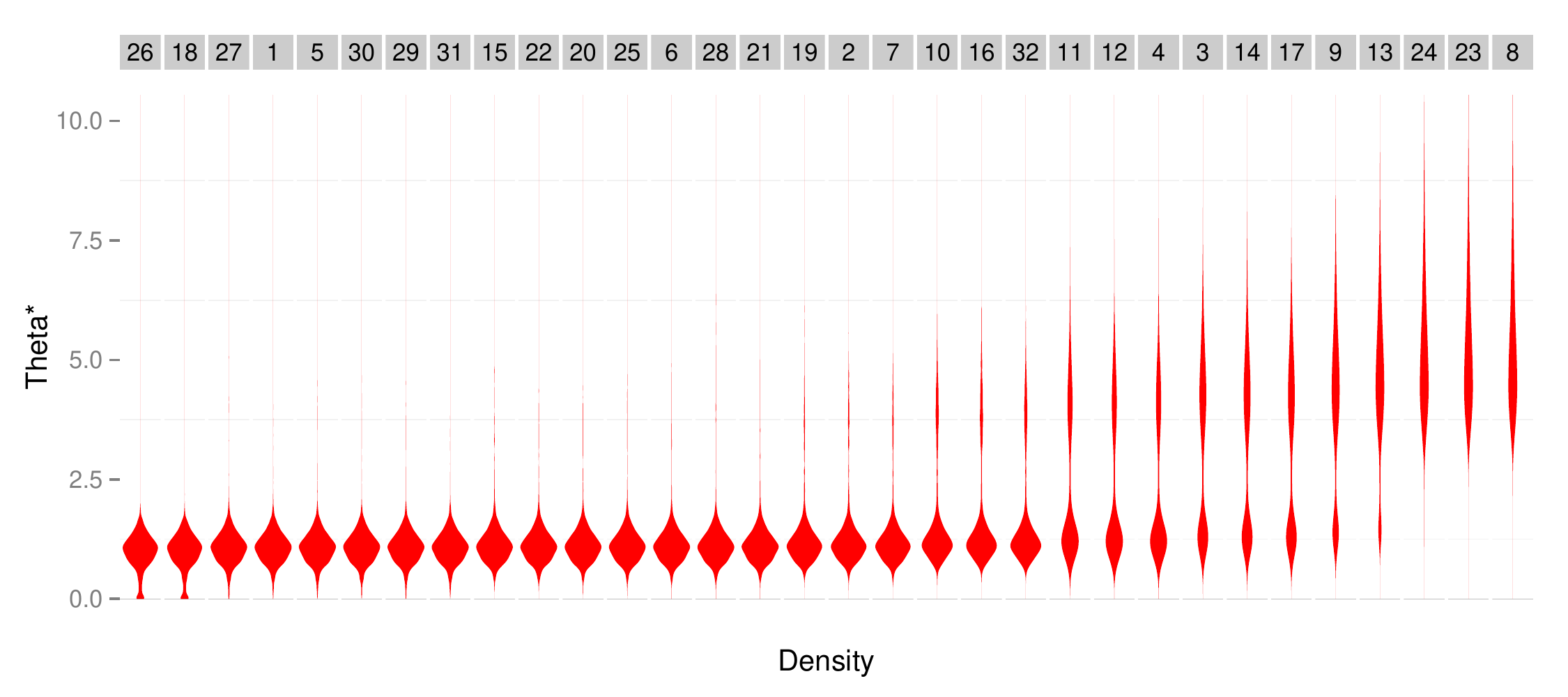}
\par\end{centering}

\protect\caption{\label{fig:Kmar-BP-thetas-1}Posterior distributions for mutation rates at each locus in \emph{K.
marmoratus} (BP population). For each distribution. the locus name is
indicated in the grey shaded box.}
\end{figure*}

\begin{figure*}[H!]
\includegraphics[width=1\textwidth]{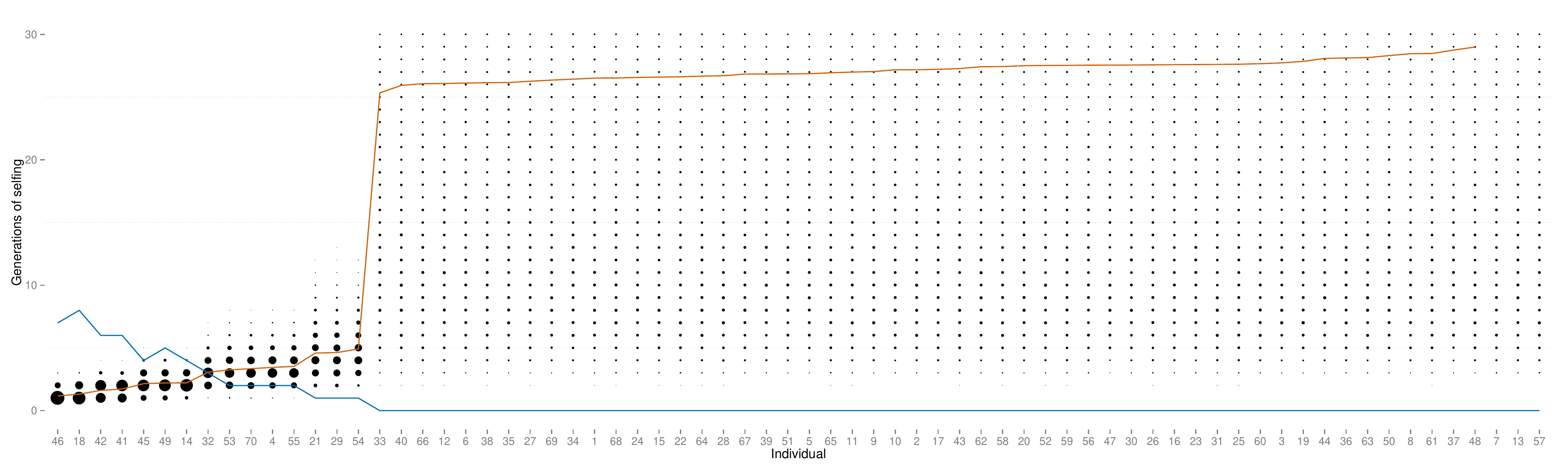}

\protect\caption{\label{fig:Kmar-BP-t-dist-1}Number of generations since the most recent outcross event in the ancestry of each individual in the sample from the BP population of \emph{K. marmoratus}. The area of each
dot indicates the posterior probability that an individual (X-axis) has the indicated number (Y-axis) of consecutive generations of selfing in its immediate ancestry.
The blue line indicates the posterior mean number
of selfing generations and the red line indicates the number of heterozygous
loci across individuals. The Y-axis is truncated to $[0,30]$.}
\end{figure*}

Figure \ref{fig:Kmar-BP-t-dist-1} shows the posterior distributions of number of generations since the most recent outcross event \eqref{eq:tlist}.

\begin{figure*}[h!]
\begin{centering}
\includegraphics[width=1\textwidth]{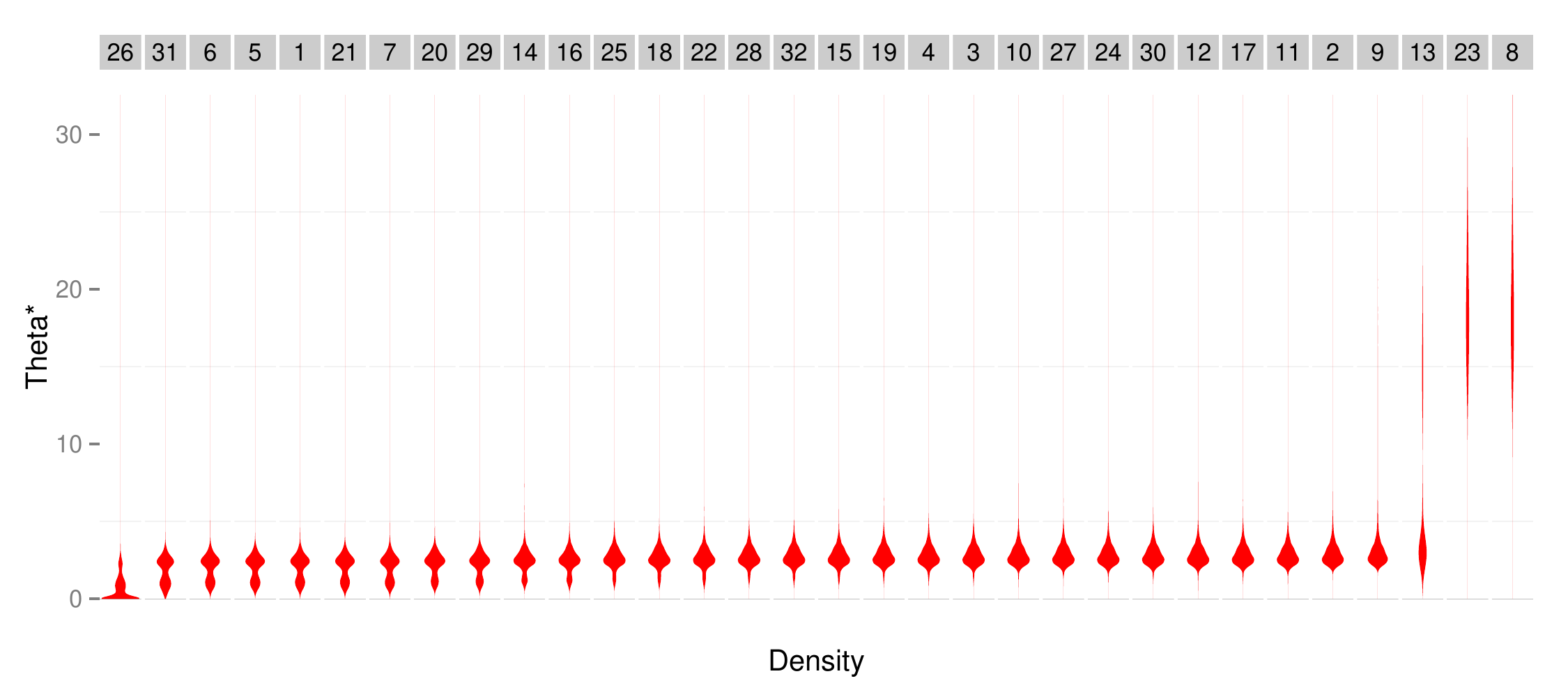}
\par\end{centering}

\protect\caption{\label{fig:Kmar-PG-thetas-1-1}Mutation rates at each locus for \emph{K.
marmoratus} (TC population).  For each distribution. the locus name is
indicated in the grey shaded box.}
\end{figure*}

\begin{figure*}
\includegraphics[width=1\textwidth]{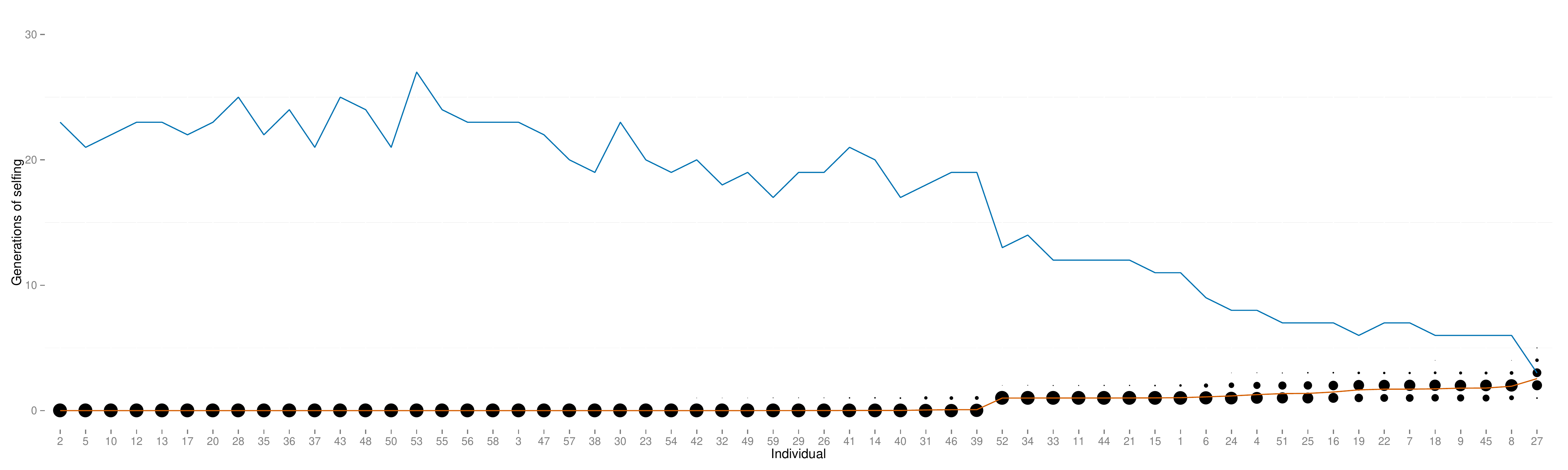}

\protect\caption{\label{fig:Kmar-PG-t-dist-1-1}Number of generations since the most recent outcross event in the ancestry of each individual in the sample from the TC population of \emph{K. marmoratus}.  Symbols as in Figure \ref{fig:Kmar-BP-t-dist-1}.}
\end{figure*}

\paragraph{Higher outcrossing rate:} Figure \ref{fig:Kmar-PG-thetas-1-1} presents posterior distributions of locus-specific mutation rates (compare Figure \ref{fig:Kmar-BP-thetas-1}). 
For each individual in the TC sample, Figure \ref{fig:Kmar-PG-t-dist-1-1} shows the posterior distribution of the number of consecutive generations of selfing in its immediate ancestry.

\subsection{Gynodioecious plant}

\begin{figure}[h!]
\begin{centering}
\includegraphics[width=0.49\textwidth]{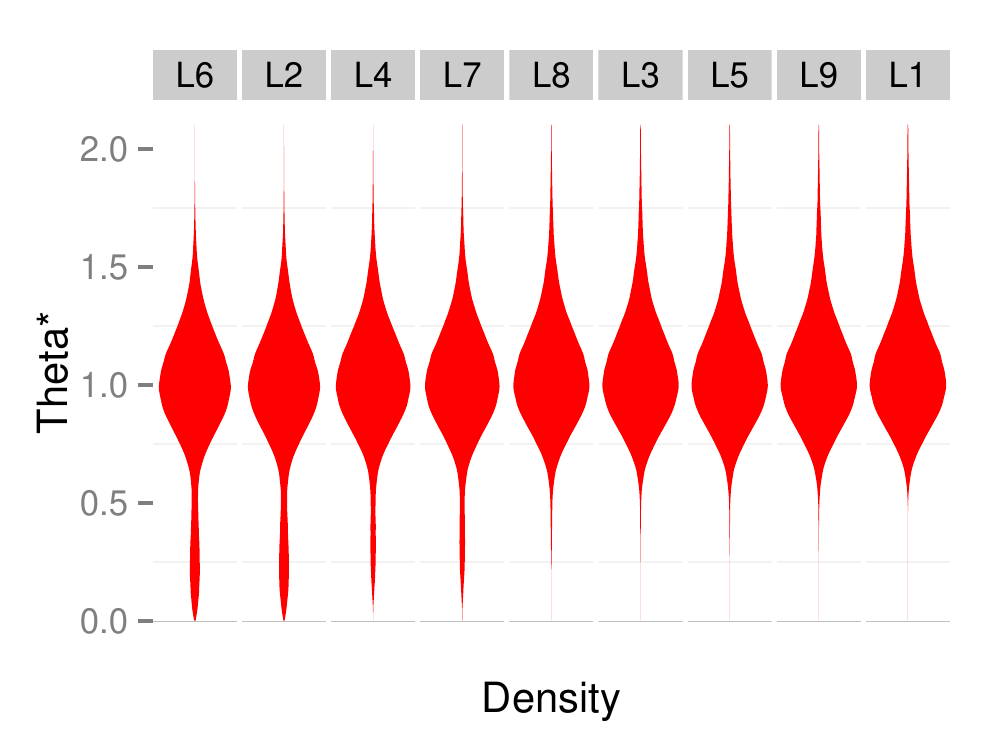}
\par\end{centering}

\protect\caption{\label{fig:Schiedea-thetas-1}
Posterior distributions
  for mutation rates at locus in \emph{S. salicaria}. For each distribution, the locus name is
indicated in the grey shaded box.
}
\end{figure}

Figure \ref{fig:Schiedea-thetas-1} presents posterior distributions for locus-specific mutation rates inferred from the  \emph{S. salicaria} data set.  The loci appear to have similar posterior medians.

\begin{figure*}
\begin{centering}
\includegraphics[width=0.49\textwidth]{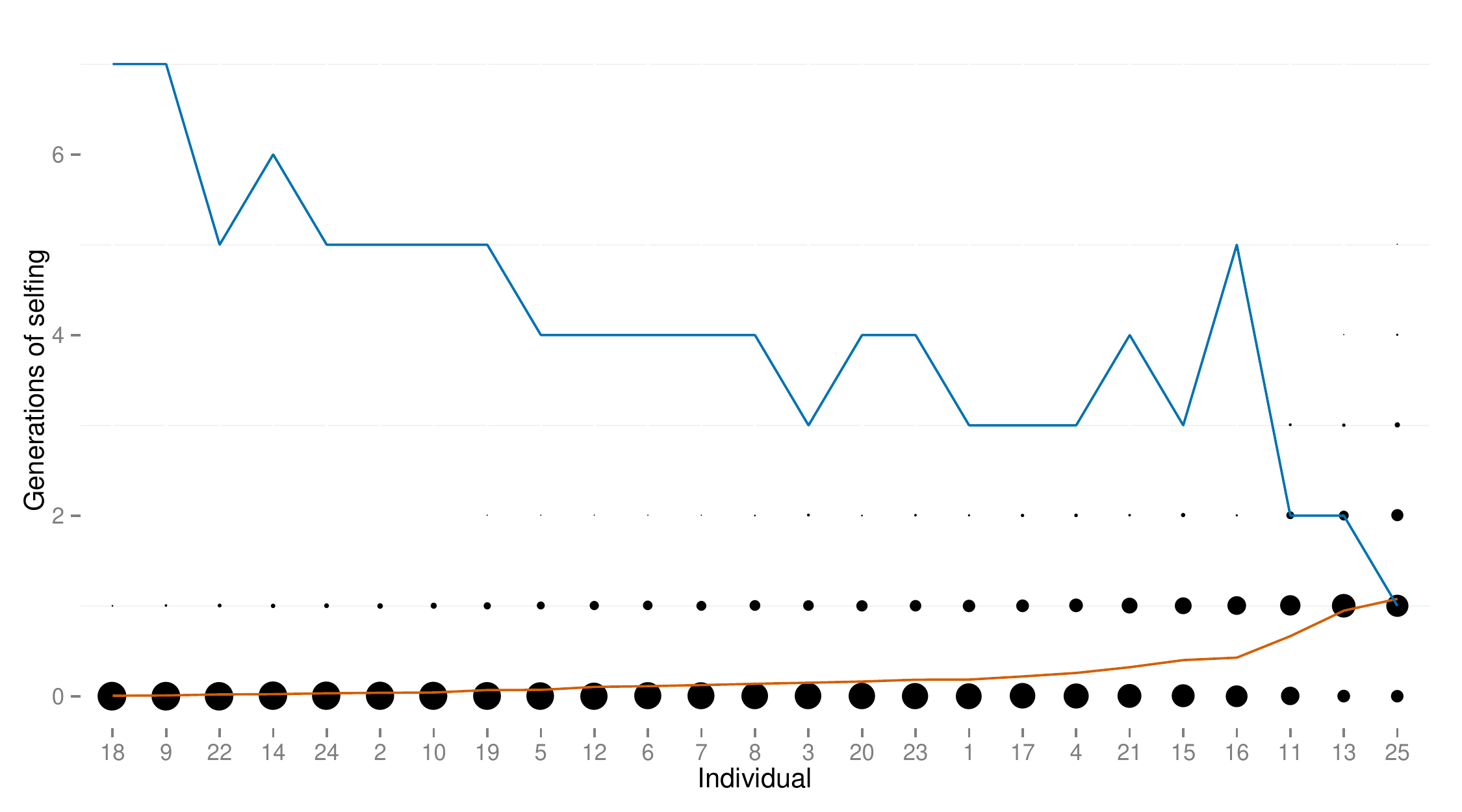}
\par\end{centering}

\protect\caption{\label{fig:Schiedea-t-dist-1}Estimated number of selfing generations
for each individual for \emph{S. salicaria}. The area of each dot indicates the
posterior probability that a numbered individual (x-axis) has been
selfed for a given number of generations (y-axis).  For each individual
the blue line indicates the posterior mean number of selfing generations
and the red line indicates the number of heterozygous loci.}
\end{figure*}

Figure \ref{fig:Schiedea-t-dist-1}
presents the inferred number of generations since the most recent outcross event $T_k$ \eqref{eq:tlist} for each individual $k$.

\begin{figure*}
\subfloat[]{\begin{centering}
\includegraphics[width=0.48\textwidth]{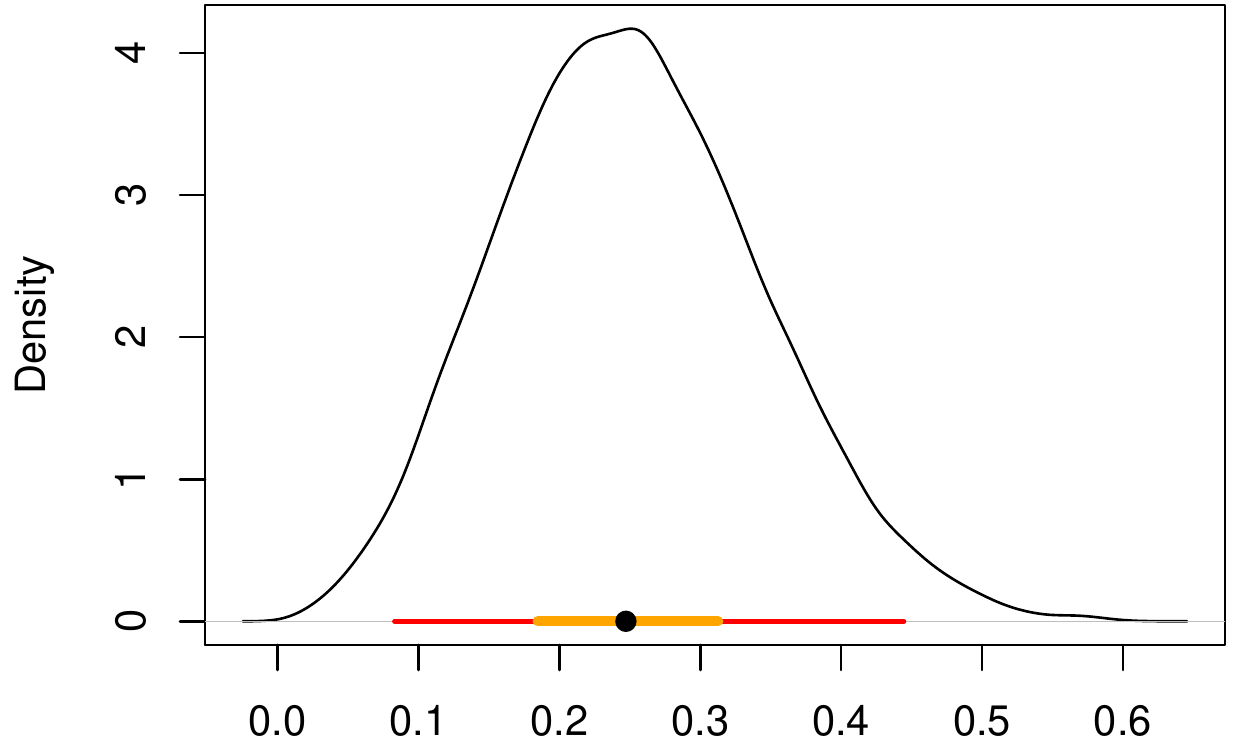}
\par\end{centering}

}\subfloat[]{\begin{centering}
\includegraphics[width=0.48\textwidth]{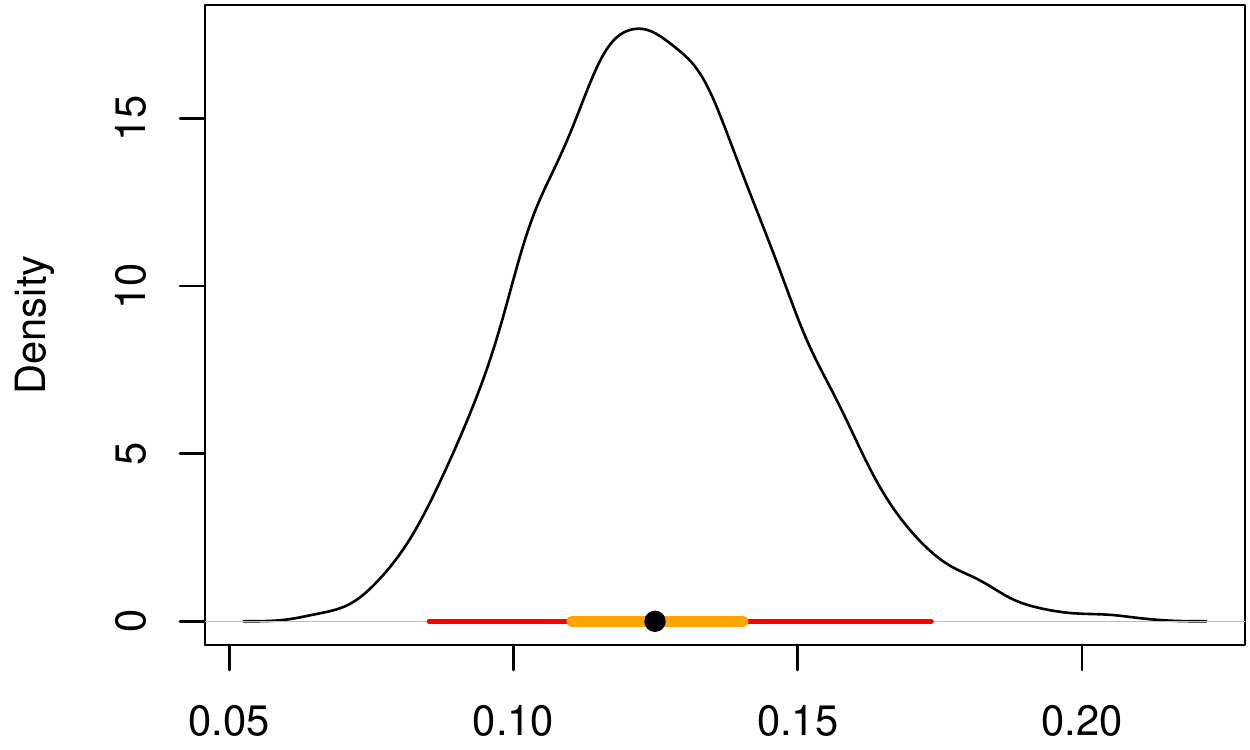}
\par\end{centering}

}

\subfloat[]{\begin{centering}
\includegraphics[width=0.48\textwidth]{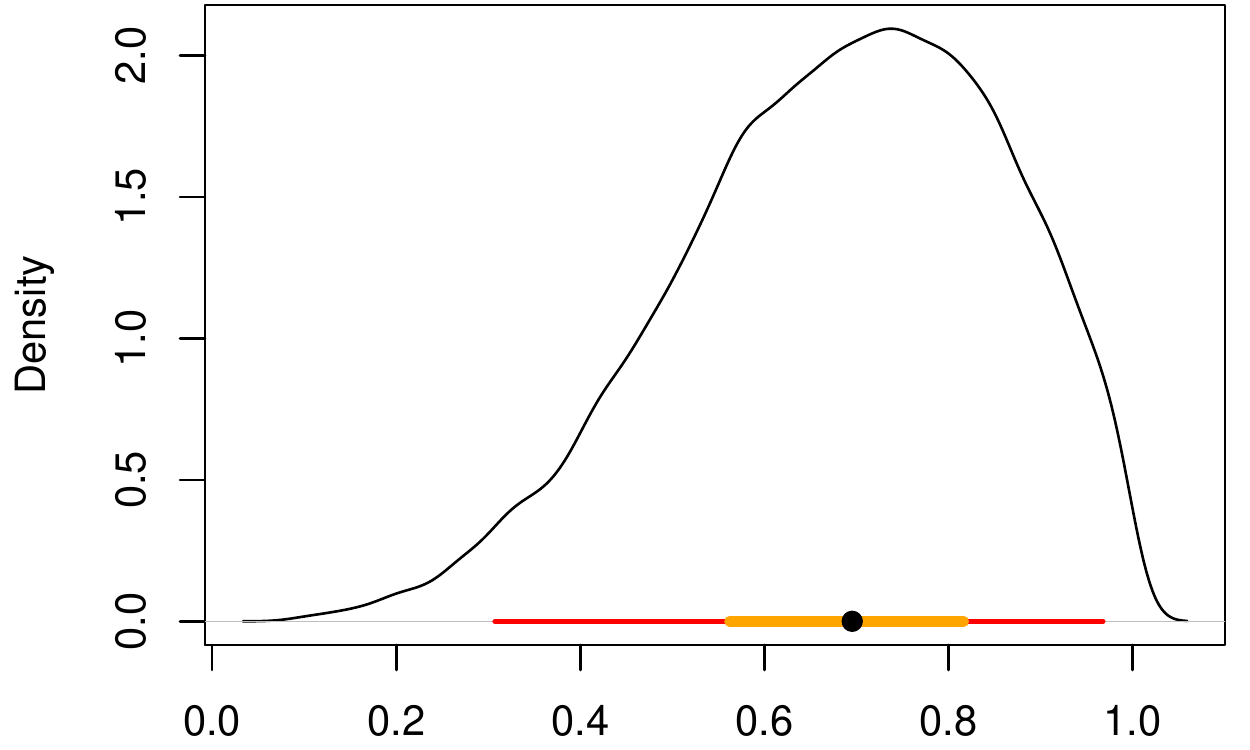}
\par\end{centering}

}\subfloat[]{\begin{centering}
\includegraphics[width=0.48\textwidth]{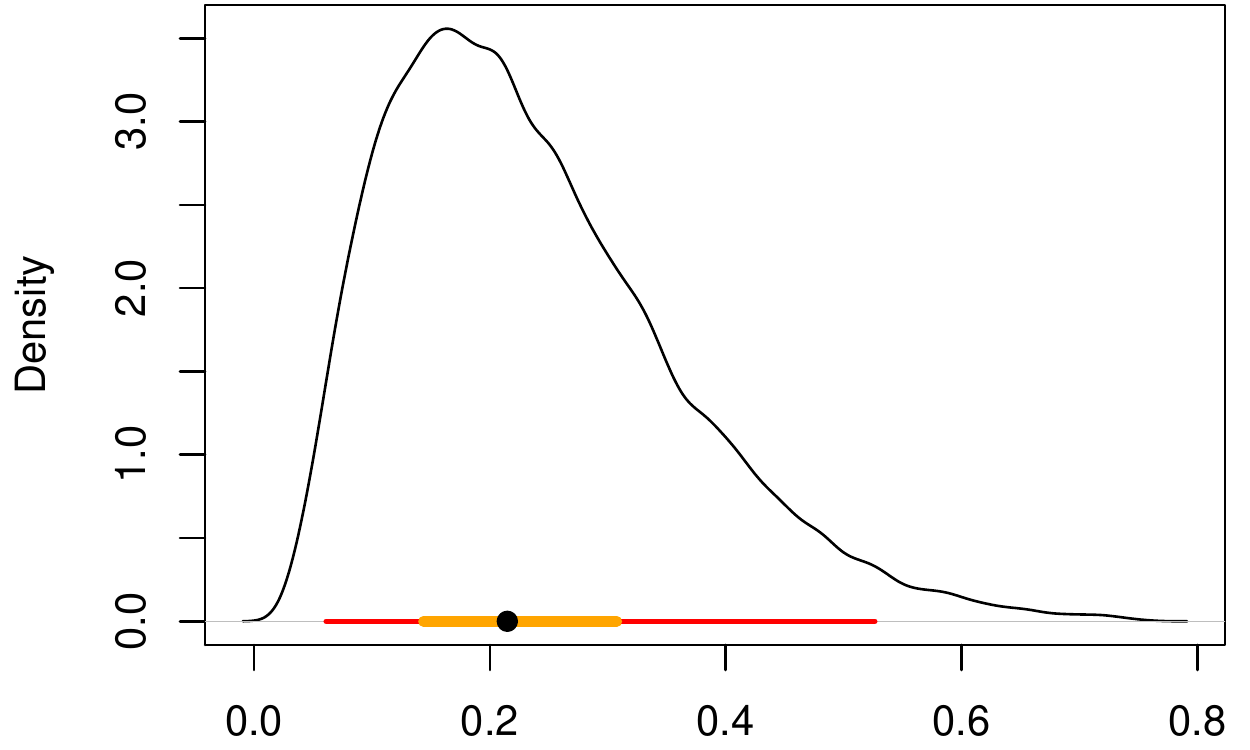}
\par\end{centering}

}

\protect\caption{\label{fig:Posterior-distributions-Schiedea}Posterior distributions
on (a) $s_G$, (b) $p_{f}$, (c) $a$, and (d) $\tau$ for the \emph{Schiedea
salicaria} data set. Also shown are 95\% BCI (red), 50\% BCI (orange), and
median (black dot).}
\end{figure*}

Figure \ref{fig:Posterior-distributions-Schiedea} presents posterior distributions for the uniparental proportion  ($s_G$), the proportion of females among reproductives ($p_{f}$), the proportion of seeds set by hermaphrodites by self-pollen ($a$), and the  viability of uniparental offspring relative to biparental offspring ($\tau$).

\end{document}